\documentclass[twocolumn,prb,showpacs,superscriptaddress]{revtex4}  

\usepackage[draft]{hyperref} 
\usepackage{graphicx}%
\usepackage{dcolumn}
\usepackage{amsmath}

\usepackage{color}
\begin{document}

\title{Terahertz time-domain spectroscopy of electromagnons in multiferroic perovskite manganites
}

\author{N. Kida}\email{n-kida@erato-mf.t.u-tokyo.ac.jp}

\affiliation{Multiferroics Project (MF), ERATO, Japan Science and Technology Agency (JST), c/o Department of Applied Physics, The University of Tokyo, 7-3-1 Hongo, Bunkyo-ku, Tokyo 113-8656, Japan}

\author{Y. Takahashi}
\affiliation{Multiferroics Project (MF), ERATO, Japan Science and Technology Agency (JST), c/o Department of Applied Physics, The University of Tokyo, 7-3-1 Hongo, Bunkyo-ku, Tokyo 113-8656, Japan}

\author{J. S. Lee}
\affiliation{Multiferroics Project (MF), ERATO, Japan Science and Technology Agency (JST), c/o Department of Applied Physics, The University of Tokyo, 7-3-1 Hongo, Bunkyo-ku, Tokyo 113-8656, Japan}

\author{R. Shimano}
\affiliation{Multiferroics Project (MF), ERATO, Japan Science and Technology Agency (JST), c/o Department of Applied Physics, The University of Tokyo, 7-3-1 Hongo, Bunkyo-ku, Tokyo 113-8656, Japan}
\affiliation{Department of Physics, The University of Tokyo, 7-3-1 Hongo, Bunkyo-ku, Tokyo 113-0033, Japan}

\author{Y. Yamasaki}
\affiliation{Department of Applied Physics, The University of Tokyo, 7-3-1 Hongo, Bunkyo-ku, Tokyo 113-8656, Japan}

\author{Y. Kaneko}
\affiliation{Multiferroics Project (MF), ERATO, Japan Science and Technology Agency (JST), c/o Department of Applied Physics, The University of Tokyo, 7-3-1 Hongo, Bunkyo-ku, Tokyo 113-8656, Japan}

\author{S. Miyahara}
\affiliation{Multiferroics Project (MF), ERATO, Japan Science and Technology Agency (JST), c/o Department of Applied Physics, The University of Tokyo, 7-3-1 Hongo, Bunkyo-ku, Tokyo 113-8656, Japan}

\author{N. Furukawa}
\affiliation{Multiferroics Project (MF), ERATO, Japan Science and Technology Agency (JST), c/o Department of Applied Physics, The University of Tokyo, 7-3-1 Hongo, Bunkyo-ku, Tokyo 113-8656, Japan}
\affiliation{Department of Physics and Mathematics, Aoyama Gakuin University, 5-10-1 Fuchinobe, Sagamihara, Kanagawa 229-8558, Japan}

\author{T. Arima}
\affiliation{Institute of Multidisciplinary Research for Advanced Materials, Tohoku University, 2-1-1 Katahira, Aoba-ku, Sendai 980-8577, Japan}

\author{Y. Tokura}
\affiliation{Multiferroics Project (MF), ERATO, Japan Science and Technology Agency (JST), c/o Department of Applied Physics, The University of Tokyo, 7-3-1 Hongo, Bunkyo-ku, Tokyo 113-8656, Japan}
\affiliation{Department of Applied Physics, The University of Tokyo, 7-3-1 Hongo, Bunkyo-ku, Tokyo 113-8656, Japan}
\affiliation{Cross-Correlated Materials Research Group (CMRG), Advanced Science Institute, RIKEN, 2-1 Hirosawa, Wako 351-0198, Japan}

\begin{abstract}Recent spectroscopic studies at terahertz frequencies for a variety of multiferroics endowed with both ferroelectric and magnetic orders have revealed the possible emergence of a new collective excitation, frequently referred to as electromagnon. It is magnetic origin, but becomes active in response to the electric field component of light. Here we give an overview on our recent advance in the terahertz time-domain spectroscopy of electromagnons or electric-dipole active magnetic resonances, focused on perovskite manganites---$R$MnO$_3$ ($R$ denotes rare-earth ions). The respective electric and magnetic contributions to the observed magnetic resonance are firmly identified by the measurements of the light-polarization dependence using a complete set of the crystal orientations. We extract general optical features in a variety of the spin ordered phases, including the $A$-type antiferromagnetic, collinear spin ordered, and ferroelectric $bc$ and $ab$ spiral spin ordered phases, which are realized by tuning the chemical composition of $R$, temperature, and external magnetic field. In addition to the antiferromagnetic resonances of Mn ions driven by the magnetic field component of light, we clarify that the electromagnon appears only for light polarized along the $a$-axis even in the collinear spin ordered phase and grows in intensity with evolution of the spiral spin order, but independent of the direction of the spiral spin plane ($bc$ or $ab$) or equivalently the direction of the ferroelectric polarization $P_{\rm s}$ ($P_{\rm s}\| c$ or $P_{\rm s}\| a$). A possible origin of the observed magnetic resonances at terahertz frequencies is discussed by comparing the systematic experimental data presented here with theoretical considerations based on Heisenberg model.
\end{abstract}


\date{Sub. 15 Mar. 2009; Invited Paper; Accepted in Special Issue on ``Terahertz Wave Photonics" in J. Opt. Soc. Am. B}

\maketitle

\section{INTRODUCTION}

Recently, there is growing interest in the magnetoelectric (ME) effect of oxide compounds endowed with both ferroelectricity and magnetism, termed multiferroics \cite{YTokura1}. The ME effect as a generic characteristic of the multiferroics is manifested by the magnetic field $H$ control of the ferroelectric polarization $P_{\rm s}$ \cite{TKimura1} or inversely the electric field control of the magnetization \cite{TLottermoser}. Such versatile cross-correlation phenomena are highly non-trivial, but can host the demanding functionality in the future electronics; this is the reason why the study of the multiferroics in a form of bulk as well as thin film is rapidly accelerated \cite{YTokura1,SWCheong,RRamesh,YTokura2}. This recent boom is triggered by the discovery of the ferroelectricity in orthorhombically distorted perovskite manganites, $R$MnO$_3$, where $R$ represents rare-earth ions such as Gd, Tb, Dy, and their solid solutions, by Kimura {\it et al.} \cite{TKimura1,TGoto1,TKimura2}. Noticeably, the direction of $P_{\rm s}$ can be flopped from the $c$- to $a$-axis by an application of the external $H$ along the $a$- or $b$-axis \cite{TKimura1,TGoto1,TKimura2}. Within the framework of Landau theory, the non-zero component of the ME tensor can produce the linear ME effect, which has been known to emerge in noncentrosymmetric magnets such as Cr$_2$O$_3$, in which both space-inversion and time-reversal symmetries are simultaneously broken \cite{WEerenstein,THODell,AFreeman}. However, the gigantic ME effect observed in $R$MnO$_3$ is better understood in terms of the phase transition between the multiferroics states; this requires the new mechanism of the ferroelectricity in these compounds and a variety of experimental approaches have been taken to reveal the multiferroic nature of $R$MnO$_3$ \cite{YTokura1,SWCheong,RRamesh,YTokura2}.

As a typical example of the ferroelectric behavior in $R$MnO$_3$, we show in {Figs. \ref{TbMnO3-Ps-epsion-T}(a) and  \ref{TbMnO3-Ps-epsion-T}(b) the temperature dependence of dielectric constant $\epsilon$ at 10 kHz and $P_{\rm s}$ along each crystallographic axis of TbMnO$_3$ in zero $H$, respectively \cite{TKimura1,TGoto1,TKimura2}. Below the N\'{e}el transition temperature $T_{\rm N}$ of 39 K, where the collinear spins of Mn ions sinusoidally order along the $b$-axis [see, the schematic illustration of the collinear spin order in Fig. \ref{RMnO3phasediagram}(c)], $\epsilon$ along the $a$-axis tends to slightly increase and shows a kink at 28 K. Below 28 K, $P_{\rm s}$ along the $c$-axis steeply increases and reaches the maximum ($\sim800$ $\mu$C/m$^2$) at the lowest temperature. Accordingly, $\epsilon$ along the $c$-axis exhibits a sharp peak at 28 K, signaling the ferroelectric phase transition. In this ferroelectric phase, the presence of the $bc$ spiral spin structure, where the spins of Mn ions rotate within the $bc$ plane with the propagation vector (along the $b$-axis), was identified by neutron scattering experiments [see, the schematic illustration of the $bc$ spiral spin order in Fig. \ref{RMnO3phasediagram}(e)] \cite{MKenzelmann}.

\begin{figure}[bt]
\includegraphics[width=0.38\textwidth]{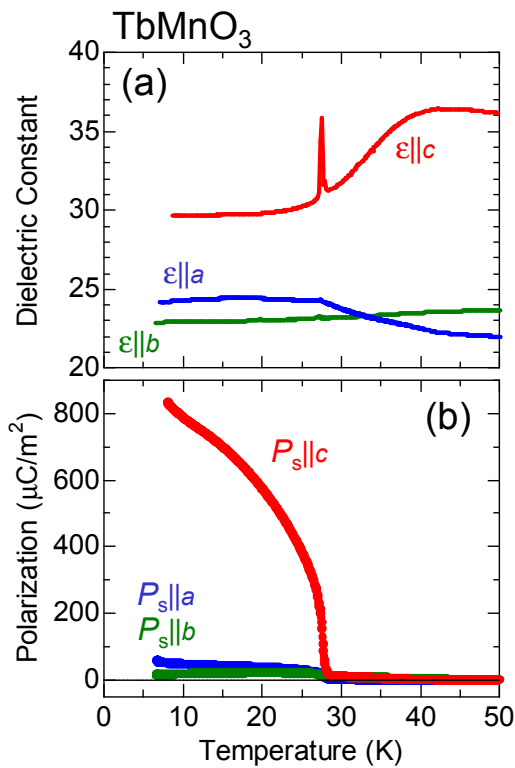}
\caption{(Color online) Temperature dependence of (a) dielectric constant $\epsilon$ at 10 kHz and (b) ferroelectric polarization $P_{\rm s}$ of TbMnO$_3$ with each crystallographic axis in zero magnetic field \cite{TKimura2}. }
\label{TbMnO3-Ps-epsion-T}
\end{figure}

As an origin of such a magnetically driven ferroelectricity in multiferroics, Katsura, Nagaosa, and Balatsky have proposed the spin-current model \cite{HKatsura1}, that $P_{\rm s}$ can be produced by the non-collinear spin order. In this model, the ferroelectricity shows up along the direction perpendicular to the wave vector and within the spiral spin plane, as given by
\begin{equation}
P_{\rm s}\propto e_{ij}\times(S_i\times S_j),
\label{spincurrent}
\end{equation}
where $e_{ij}$ is a unit vector connecting the nearest-neighbor spins, $S_i$ and $S_j$, as shown in Fig. \ref{RMnO3phasediagram}(e). According to Eq. (\ref{spincurrent}), $P_{\rm s}$ is expected to emerge along the $c$-axis in $R$MnO$_3$ with the $bc$ spiral spin order, being consistent with the experimental observation of TbMnO$_3$ shown in Fig. \ref{TbMnO3-Ps-epsion-T}. Furthermore, the relationship between the ferroelectricity and the spin structure could be directly revealed in the spin-polarized neutron scattering by controlling the vector chirality defined by $(S_i\times S_j)$ of TbMnO$_3$ by an external electric field \cite{YYamasaki1}. Therefore, the observed $P_{\rm s}$-flop from the $c$- to $a$-axis by applying the external $H$ can be regarded as the flop of the spiral spin plane from $bc$ to $ab$. The presence of the $ab$ spiral spin order with $P_{\rm s}$ along the $a$-axis was recently confirmed for TbMnO$_3$ in $H$ along the $b$-axis \cite{NAliouane} and Gd$_{0.7}$Tb$_{0.3}$MnO$_3$ in zero $H$ \cite{YYamasaki2}. The equivalent expression to the spin-current model has also been obtained phenomenologically \cite{MMostovoy} and by considering the spin-lattice coupling through the Dzyaloshinski-Moriya interaction \cite{IASergienko}. Recently, ME phase diagrams of $R$MnO$_3$ were theoretically investigated on the basis of numerical calculations of a microscopic spin model including the anisotropy terms and the Dzyaloshinski-Moriya interaction, which successfully reproduces the experimental phase diagrams including flops, emergence, and disappearance of $P_{\rm s}$ \cite{MMochizuki}.

The ME coupling between ferroelectricity and magnetism would also produce the intriguing lower-lying spin excitation \cite{GASmolenski}. Since 1960s, the presence of such an excitation was theoretically inferred in the ME materials \cite{VGBaryakhtar}. This new elementary excitation appears as a magnetic resonance in the dielectric constant spectrum $\epsilon(\omega)$ as a response to the electric field component of light $E^\omega$, thus can be termed the electric-dipole active magnetic resonance. This is in contrast to the case of the magnetic resonance appearing in the magnetic permeability spectrum $\mu(\omega)$ as a result of the spin excitation driven by the magnetic field component of light $H^\omega$ \cite{AMBalbashov}. Recently, Pimenov {\it et al.} have measured the optical spectra of TbMnO$_3$ and GdMnO$_3$ in the energy range of 0.4--4.8 meV by using a backward wave oscillator (BWO) as a light source, combined with Mach-Zehnder interferometry, in which the complex optical constants can be extracted in the quasi-optical setup \cite{APimenov1}. They found a single peak-structure in the imaginary part of $\epsilon(\omega)$, at 2.9 meV for TbMnO$_3$ and 2.5 meV for GdMnO$_3$ around 10 K in zero $H$. These absorptions were shown to be allowed when $E^\omega$ was set parallel to the $a$-axis. Furthermore, the dramatic reduction of the real part of $\epsilon(\omega)$ as a result of the transformation from the $bc$ spiral spin order to the $A$-type AFM order, was observed by applying the external $H$ along the $c$-axis. Based on these facts, they claimed the possible emergence of the electric-dipole active magnetic resonances in these compounds, now frequently refereed to as electromagnons. This pioneering work stimulates the variety of investigations concerning the low-energy spin dynamics of multiferroics such as perovskite $R$MnO$_3$ \cite{APimenov1,HKatsura2,DSenff1,APimenov2,RValdesAguilar1,APimenov3,NKida1,DSenff2,YTakahashi1,NKida2,NKida3,RValdesAguilar2,SMiyahara,APimenov4,JSLee,YTakahashi2,JSLee2,APimenovR,DSenffR}, hexagonal YMnO$_3$ \cite{SPailhes}, BiFeO$_3$ \cite{MCazayous,MKSingh}, and $R$Mn$_2$O$_5$ \cite{EGolovenchits,ABSushkov,CFang}, by using far-infrared optical, Raman scattering, and inelastic neutron scattering spectroscopies.

Within the framework of the spin-current model that can explain the ferroelectric properties observed in $R$MnO$_3$, Katsura, Balatsky, and N. Nagaosa showed that $E^\omega$ is possible to drive the oscillation of the spiral spin plane in $R$MnO$_3$, which produces the magnetic resonance at terahertz frequencies along the direction perpendicular to the spiral spin plane \cite{HKatsura2}; $E^\omega \| a$ and $E^\omega \| c$ in the $bc$ and $ab$ spiral spin ordered phases, respectively. This collective mode can be termed the rotation mode of the spiral spin plane, as the example of the manifestation of the dynamical ME coupling in multiferroics. Subsequently, Senff {\it et al.} performed inelastic neutron scattering experiments of TbMnO$_3$ in the $bc$ spiral spin ordered phase and found the magnetic excitation around 2 meV at $k=0$ \cite{DSenff1}, the peak position of which is nearly identical to that of the electromagnon in TbMnO$_3$ revealed by BWO spectroscopy \cite{APimenov1}. Furthermore, they also clarified that the observed mode is polarized along the $a$-axis perpendicular to the spiral spin plane \cite{DSenff1}, as predicted by theoretical considerations. Therefore, the electromagnon observed in $R$MnO$_3$ at terahertz frequencies was anticipated to be ascribed to the rotation mode of the spiral spin plane.

Here we show our latest advance in the terahertz time-domain spectroscopy on electromagnons or electric-dipole active magnetic resonances in multiferroic perovskite manganites $R$MnO$_3$ \cite{NKida1,YTakahashi1,NKida2,NKida3,SMiyahara}. Contrary to early assignments described above, we provide the compelling evidence that the electromagnon observed in $R$MnO$_3$ cannot be ascribed to the rotation mode of the spiral spin plane; it appears only for $E^\omega \| a$, irrespective of the direction of the spiral spin plane ($bc$ or $ab$). This conclusion is unambiguously revealed by the systematic optical investigations on the spin excitation at terahertz frequencies in $R$MnO$_3$ based on light-polarization, temperature, and external $H$ dependence. We clarified the full spectral shape of the electromagnon, which consists of two peak-structures around 2 meV and 5--8 meV. The lower-lying electromagnon survives even in the collinear spin ordered phase. Below $T_{\rm N}$, we also identify the presence of the antiferromagnetic resonances of Mn ions around 2 meV driven by $H^\omega \| a$ or $H^\omega \| c$, the peak position of which is nearly identical to that of the lower-lying electromagnon for $E^\omega \| a$.

The format of this article is as follows. In Sec. \ref{Methods}, we briefly describe the experimental setup for the terahertz time-domain spectroscopy we used here and the estimate procedure of the complex optical constants from the raw data in time domain. Sec. \ref{Results} is devoted to show the results of the optical spectra at terahertz frequencies for a variety of spin ordered phases of $R$MnO$_3$, as tuned by the ionic radius of $R$, temperature, and external $H$. After the brief descriptions of the overall optical spectrum of TbMnO$_3$ (Sec. \ref{ExpTbMnO3}) and the basic feature of $R$MnO$_3$ in terms of Mn-O-Mn bond angle (Sec. \ref{ExpRMnO3}), we sum up the general optical properties in a variety of spin ordered phases by taking DyMnO$_3$ (Sec. \ref{ExpDyMnO3}) \cite{NKida1} and Gd$_{0.7}$Tb$_{0.3}$MnO$_3$ (Sec. \ref{ExpGd07Tb03MnO3}) \cite{NKida2} as the examples. We discuss in Sec. \ref{Discussion} the origin of the electromagnons in $R$MnO$_3$ with theoretical considerations based on the Heisenberg model \cite{SMiyahara}. Summary is given in Sec. \ref{Summary}.

\begin{figure*}[bt]
\includegraphics[width=0.88\textwidth]{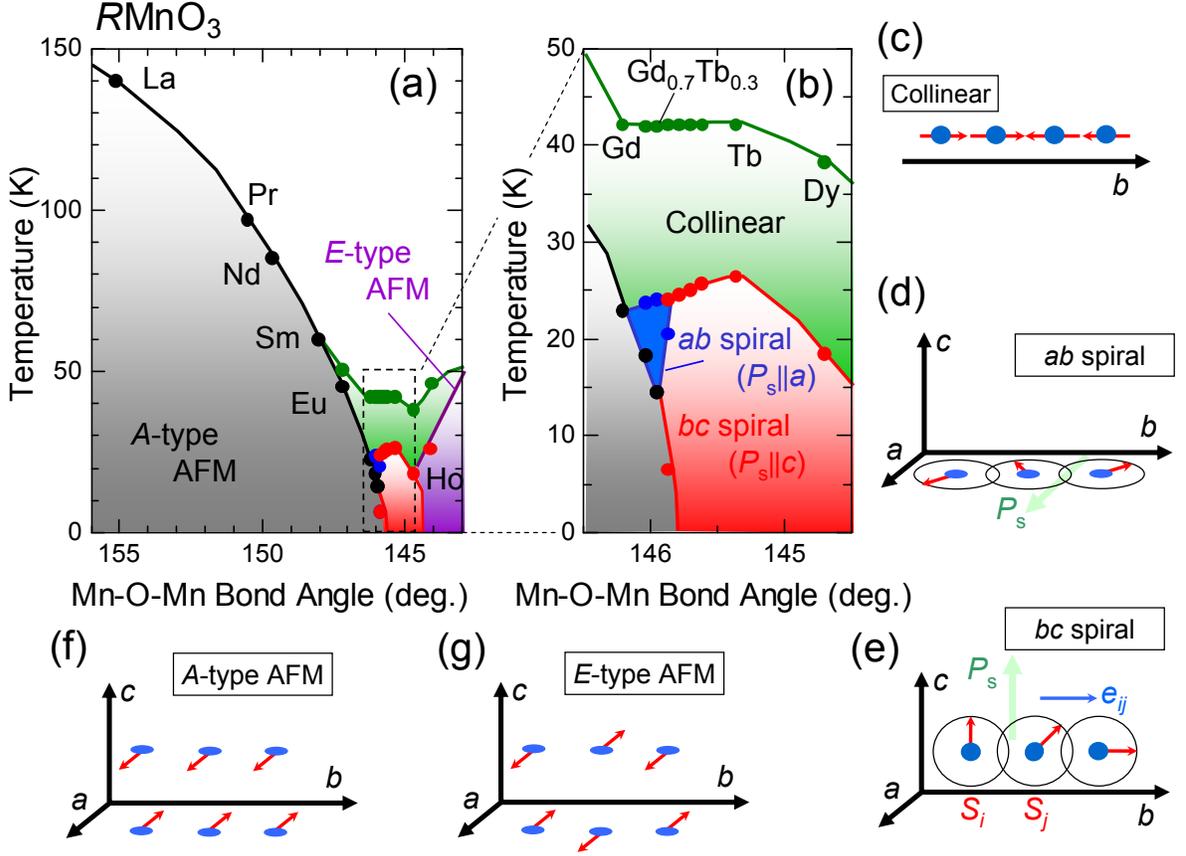}
\caption{(Color online) (a) Magnetoelectric phase diagrams of $R$MnO$_3$ with change of the Mn-O-Mn bond angle, as reproduced from Refs. \cite{TGoto2,TKimura3} ($R$ represents rare-earth ions). The solid lines are merely the guide to the eyes. (b) Magnified view of (a), showing the emergence of the ferroelectric phases for $R=$ Gd, Tb, and Dy. By changing the Mn-O-Mn bond angle, $R$MnO$_3$ shows the large variations of the spin structures with temperature, including the $A$-type (layer-type) antiferromagnetic (AFM), collinear spin ordered, $bc$ and $ab$ spiral spin ordered, and $E$-type AFM phases. The schematic illustrations of these phases are shown in (c)--(g); Mn ions and their spins are highlighted by circles and arrows, respectively. $e_{ij}$ is the unit vector connecting adjacent spins, $S_i$ and $S_j$. According to the spin-current mechanism, as formulated by Eq. (\ref{spincurrent}), the ferroelectric polarization $P_{\rm s}$ appears along the $c$- and $a$-axes in the $bc$ and $ab$ spiral spin ordered phases, respectively. }
\label{RMnO3phasediagram}
\end{figure*}

\begin{figure*}[bt]
\includegraphics[width=0.88\textwidth]{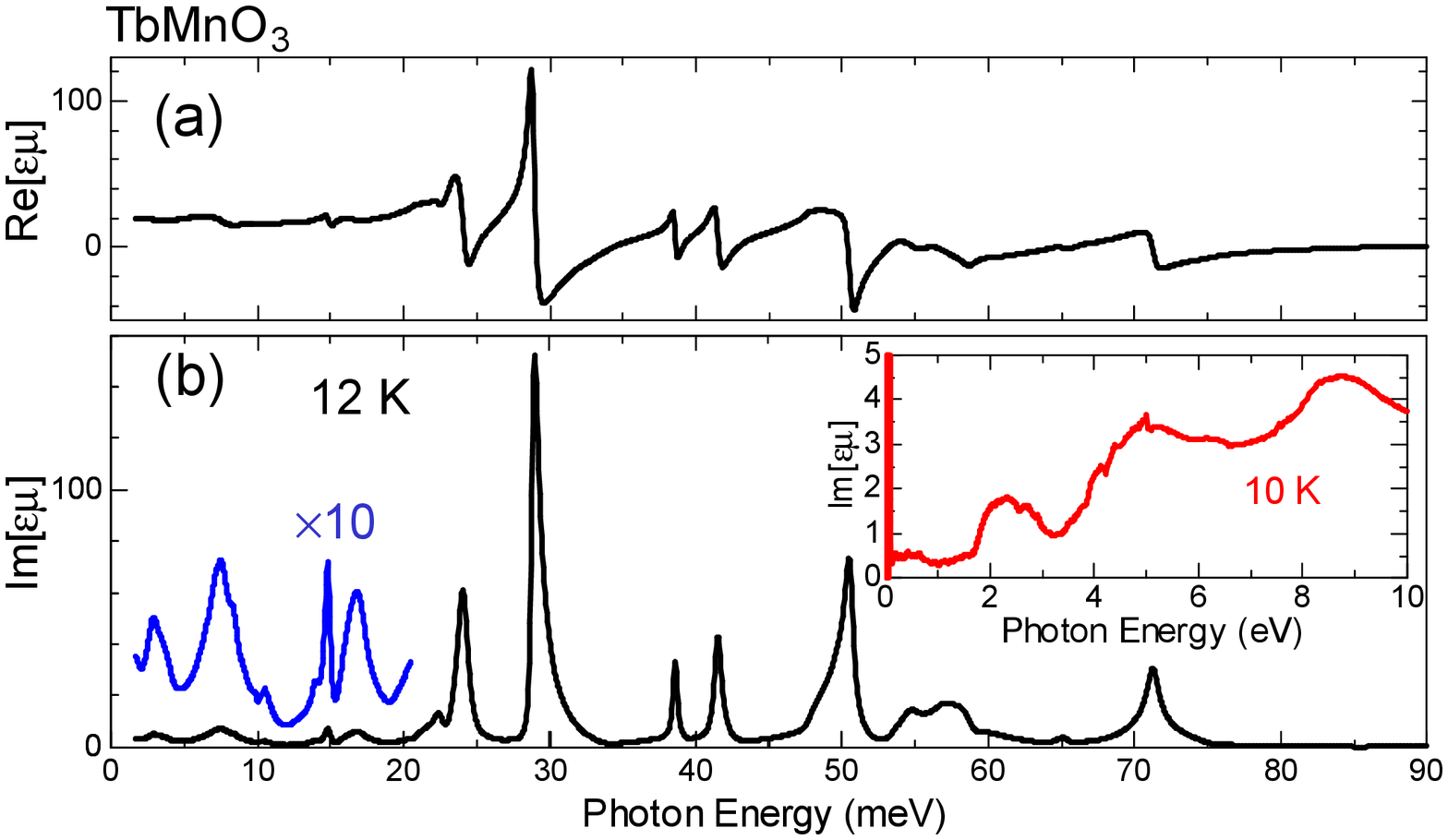}
\caption{(Color online) Overall optical spectrum of TbMnO$_3$, ranging from 2 meV to 10 eV, in the ferroelectric $bc$ spiral spin ordered phase, measured around 10 K. $E^\omega$ was set parallel to the $a$-axis. (a) Real Re$[\epsilon\mu]$ and (b) imaginary Im$[\epsilon\mu]$ parts of the $\epsilon\mu$ spectrum up to 90 meV. The low-energy part of the Im$[\epsilon\mu]$ spectrum below 21 meV is multiplied by 10. Inset shows the Im$[\epsilon\mu]$ spectrum up to 10 eV, measured at 10 K. Below 10 meV, both Re$[\epsilon\mu]$ and Im$[\epsilon\mu]$ spectra were obtained by terahertz time-domain spectroscopy. Above 10 meV, we used the Fourier transform spectrometer (0.01--0.8 eV) and the grating spectrometer (0.3--36 eV).}
\label{figTbMnO3}
\end{figure*}

\section{METHODS}\label{Methods}

We used the standard experimental setup for the terahertz time-domain spectroscopy in transmission geometry. Femtosecond laser pulses delivered from the mode-locked Ti:sapphire laser with the center wavelength of 800 nm, the pulse width of 100 fs, and the repetition rate of 80 MHz were divided into pump and trigger pulses. The pump pulses were irradiated to the ZnTe crystal or to the photoswitching device made on the low-temperature-grown GaAs (LT-GaAs) coupled with the bow-tie antenna. The detector was the another LT-GaAs coupled with the dipole antenna. Although the lower limit of the available energy range is restricted by the size of each sample, we cover the energy range of 2--9 meV and 0.8--6 meV by using ZnTe and LT-GaAs terahertz emitters, respectively. The radiated terahertz pulse was collimated and focused on the sample by a pair of the off-axis paraboloidal mirrors. For the light-polarization dependence, the wire-grid polarizer was inserted in between the off-axis paraboloidal mirrors. 

Single-crystalline samples were grown by the floating-zone method \cite{TKimura2}. Specimens with wide $ac$, $ab$, and $bc$ faces were cut from the bowl and each crystallographic axis was determined by back Laue photographs. The obtained specimens were characterized by x-ray diffraction, $\epsilon$ at 10 kHz, $P_{\rm s}$, and magnetization measurements, which were all consistent with previous reports in Refs. \cite{TKimura1,TGoto1,TKimura2}. For transmission experiments, we polished the specimens to the thickness of 100--850 $\mu$m. We carefully confirmed that there was no effect of the polishing procedure on the optical properties at terahertz frequencies of $R$MnO$_3$.

We estimated the optical constants $\tilde{n}$ of $R$MnO$_3$ without the Kramers-Kronig transformation. In $R$MnO$_3$, there are spin excitations driven by both $E^\omega$ and $H^\omega$ in the measured energy range, as we could clarify their contributions to $\tilde{n}$ mainly based on the measurements of the complete set of the light-polarization dependence (Sec. \ref{Results}). Due to the emergence of the magnetic resonances driven by $H^\omega$, $\tilde{n}$ should be precisely expressed as $\tilde{n}=\sqrt{\epsilon\mu}$ (whereas $\tilde{n}=\sqrt{\epsilon}$ for the case of non-magnets). We confirmed that the contribution of $\mu$ is negligible to the complex transmission coefficient by the numerical calculation and thus the effect of $H^\omega$ was taken into account by adopting $\tilde{n}=\sqrt{\epsilon\mu}$. Therefore, we used the quantity of $\epsilon\mu$ in this paper. Further details of our estimate procedure and the validity of this approach can be found in Ref. \cite{NKida1}.

\section{EXPERIMENTAL RESULTS}\label{Results}

\subsection{Overall optical spectrum of TbMnO$_{\textbf 3}$ from terahertz to ultraviolet frequencies}\label{ExpTbMnO3}

First, we show the overall optical spectrum of TbMnO$_3$ in the ferroelectric $bc$ spiral spin ordered phase, ranging from terahertz to ultraviolet frequencies \cite{YTakahashi1}. Figures \ref{figTbMnO3}(a) and \ref{figTbMnO3}(b) show the real Re$[\epsilon\mu]$ and imaginary Im$[\epsilon\mu]$ parts of the $\epsilon\mu$ spectra of TbMnO$_3$ up to 90 meV, respectively, measured at 12 K. $E^\omega$ was set parallel to the $a$-axis. Although there is no contribution of $\mu$ to the $\epsilon\mu$ spectrum above 8 meV, we used the notation of Re$[\epsilon\mu]$ and Im$[\epsilon\mu]$. We obtained the $\epsilon\mu$ spectrum above 10 meV by using the Kramers-Kronig transformation; the polarized reflectivity spectrum was measured in the energy ranges of 0.01--0.8 eV and 0.6--36 eV by using Fourier transform infrared spectrometer and grating monochromator, respectively. In the energy range of 10--22 meV, we performed both the transmission and reflectance measurements and thus directly estimated the $\epsilon\mu$ spectrum without the Kramers-Kronig transformation. Below 10 meV, terahertz time-domain spectroscopy was used in transmission geometry. The inset of Fig. \ref{figTbMnO3}(b) shows the Im$[\epsilon\mu]$ spectrum of TbMnO$_3$ up to 10 eV, measured at 10 K, where the electronic transitions are dominant. The optical transition around 2 eV across the charge-transfer gap is clearly identified. The peak-structures around 5 eV and 9 eV can be assigned to the transitions from O 2$p$ to Mn $3d$ and Tb $5d$ states, respectively, according to the systematic optical study of the transition metal oxides with perovskite structure \cite{TArima2,TArima3}. Although the GdFeO$_3$-type distortion results in the splitting and mixing of the phonon modes, we can roughly classify the character of the observed phonon modes into three types---stretching, bending, and external modes, which are typical for ideal cubic perovskite structure \cite{TArima3,ISSmirnova}. The optical phonon around 70 meV is assigned to the stretching mode of Mn-O-Mn. Several modes discerned from 25 meV to 60 meV mainly come from the bending modes of Mn-O-Mn. The optical phonon modes below 25 meV can be ascribed to the external modes, which correspond to the vibrations of Tb ions. Below 20 meV, other contributions to the far-infrared spectrum are identified as four pronounced absorption peak-structures at 3 meV, 7 meV, 14 meV, and 17 meV, as clearly seen in the Im$[\epsilon\mu]$ spectrum, multiplied by 10 [Fig. \ref{figTbMnO3}(b)]. The 14 meV peak-structure corresponds to the lowest-lying external phonon mode, which shows an appreciable coupling with the lower-lying electromagnon \cite{RValdesAguilar1,YTakahashi1,JSLee2}. According to the inelastic neutron scattering experiments \cite{RKajimoto,DSenff1}, the upper edge of the magnon band of TbMnO$_3$ is located around 8 meV. Therefore, we previously assigned the peak-structure at 17 meV to the upper edge of the two-magnon band, as the energy of this peak-structure just corresponds to the twice of the magnon energy at zone edge \cite{YTakahashi1}. However, we recently clarified that this absorption peak is diminished in other $R$MnO$_3$ such as DyMnO$_3$ \cite{JSLee}, GdMnO$_3$ \cite{JSLee}, and Eu$_{1-x}$Y$_x$MnO$_3$ ($x$ represents the nominal composition) \cite{YTakahashi2}, although two other peak-structures below 10 meV are discerned. Therefore, the peak-structure at 17 meV can be ascribed to the crystal field excitation of Tb$^{3+}$ ions, contrary to our previous assignment \cite{YTakahashi1}. Further detailed studies concerning the phonon modes in $R$MnO$_3$ will be presented elsewhere \cite{YTakahashi2,JSLee2}. In the following subsections, we focus on optical properties of $R$MnO$_3$ below 10 meV.

\begin{figure*}[bt]
\includegraphics[width=0.88\textwidth]{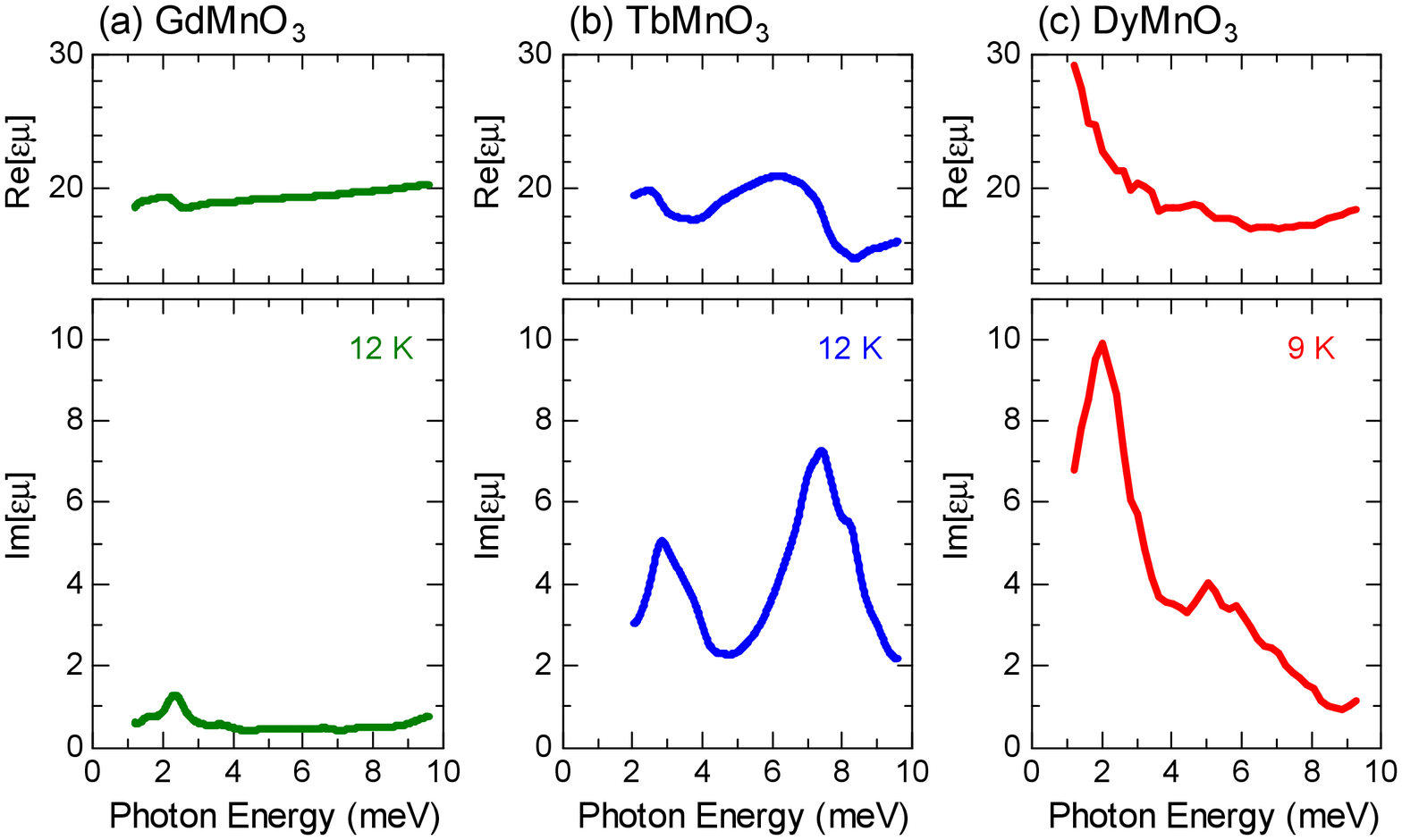}
\caption{(Color online) Low-energy electrodynamics of the spin excitation of (a) GdMnO$_3$ at 12 K, (b) TbMnO$_3$ at 12 K, and (c) DyMnO$_3$ at 9 K, for $E^\omega \| a$ and $H^\omega \| c$. Upper and lower panels show the real Re[$\epsilon\mu$] and imaginary Im[$\epsilon\mu$] parts of the $\epsilon\mu$ spectra, respectively. Around 10 K, the $bc$ spiral spin ordered phase is realized for TbMnO$_3$ and DyMnO$_3$, while the $A$-type antiferromagnetic phase for GdMnO$_3$.}
\label{figRMnO3LightP}
\end{figure*}

\subsection{Overview of terahertz spectra of $\mbox{\boldmath$R$}$MnO$_{\textbf 3}$}\label{ExpRMnO3}

Here we overview the optical spectra at terahertz frequencies in thermally induced spin ordered phases---$A$-type antiferromagnetic (AFM) and $bc$ spiral spin ordered phases, realized in a family of $R$MnO$_3$ ($R=$ Gd, Tb, and Dy). In $R$MnO$_3$, the decrease of the ionic radius of $R$ or equivalently the decrease of the Mn-O-Mn bond angle $\phi$, destabilizes the $A$-type AFM order \cite{TKimura3}, in which the Mn spins antiferromagnetically stack along the $c$-axis and ferromagnetically order along the $a$- and $b$-axes, as schematically shown in Fig. \ref{RMnO3phasediagram}(f). The ME phase diagrams of $R$MnO$_3$ are reproduced in Figs. \ref{RMnO3phasediagram}(a) and \ref{RMnO3phasediagram}(b) \cite{TKimura3,TGoto2}. LaMnO$_3$ $(\phi=155.1^\circ)$ has one $e_g$ electron per Mn-site $(d^4)$ and is an $A$-type AFM insulator below $T_{\rm N}$ of 140 K. With decreasing the ionic radius of $R$ from $R=$ La $(\phi=155.1^\circ)$ to $R=$ Gd $(\phi=146.2^\circ)$, $T_{\rm N}$ dramatically decreases from 140 K to 42 K [Fig. \ref{RMnO3phasediagram}(a)] as a result of the spin frustration caused by competing spin-exchange interactions \cite{TKimura3}. In between $R=$ Tb $(\phi=145.4^\circ)$ and $R=$ Dy $(\phi=144.7^\circ)$, the ferroelectricity emerges along the $c$-axis, accompanied by the $bc$ spiral spin order [Fig. \ref{RMnO3phasediagram}(e)] \cite{TKimura1,TGoto1,TKimura2}, as can be clearly seen in the magnified view [Fig. \ref{RMnO3phasediagram}(b)]. At 10 K, GdMnO$_3$ and Gd$_{0.7}$Tb$_{0.3}$MnO$_3$ are $A$-type AFM insulators. Noticeably, around $\phi$ of 146.0$^\circ$, as can be realized in the mixed valence Gd$_{1-x}$Tb$_x$MnO$_3$ in between the paraelectric $A$-type AFM GdMnO$_3$ and the ferroelectric TbMnO$_3$ \cite{TGoto2}, the $ab$ spiral spin ordered phase with $P_{\rm s}\| a$ [Fig. \ref{RMnO3phasediagram}(d)] emerges in a narrow range of temperature; for example, $P_{\rm s}\| a$ shows up between 16 K and 24 K for Gd$_{0.7}$Tb$_{0.3}$MnO$_3$ \cite{TGoto2}. Below 16 K, the $A$-type AFM phase is stable. For further decreasing the ionic radius of $R$, $E$-type AFM order is developed for $R=$ Ho [Fig. \ref{RMnO3phasediagram}(g)], in which the Mn spins order in the sequences of $\uparrow\uparrow\downarrow\downarrow$ along each crystallographic direction. 

We show the Re[$\epsilon\mu$] and Im[$\epsilon\mu$] spectra in upper and lower panels of Fig. \ref{figRMnO3LightP}, respectively, for (a) GdMnO$_3$ at 12 K, (b) TbMnO$_3$ at 12 K, and (c) DyMnO$_3$ at 9 K. In these measurements, $E^\omega$ and $H^\omega$ were set parallel to $a$- and $c$-axes, respectively, with use of the crystal plates with a widest $ac$ face. As can be clearly seen, there is a large variation of the optical spectra with the ionic radius of $R$. In the $A$-type AFM phase of GdMnO$_3$ $(\phi=146^\circ)$ at 12 K [Fig. \ref{figRMnO3LightP}(a)], a clear sharp peak-structure can be discerned at 2.3 meV in the Im[$\epsilon\mu$] spectrum, yielding the maximum magnitude of Im[$\epsilon\mu$] $\sim1$. In accord with this, there is a dispersive structure in the Re[$\epsilon\mu$] spectrum. This tiny absorption can be assigned to the spin excitation driven by $H^\omega \| c$, the details of which are discussed in Sec. \ref{ExpGd07Tb03MnO3} by adopting Gd$_{0.7}$Tb$_{0.3}$MnO$_3$. With decreasing the ionic radius of $R$, we can see the dramatic modification of the $\epsilon\mu$ spectra. In the $bc$ spiral spin ordered phase $(P_{\rm s}\| c)$ of TbMnO$_3$ at 12 K [Fig. \ref{figRMnO3LightP}(b)], the magnitude of Im$[\epsilon\mu]$ increases, which forms the pronounced broad continuum-like absorption composed of two peak-structures at 2.9 meV and 7.4 meV. The magnitudes of Im[$\epsilon\mu$] for the lower- and higher-lying peak-structures reach about 5 and 7, respectively. At these peak positions, we can identify the clear dispersive structures in the Re$[\epsilon\mu]$ spectrum. In DyMnO$_3$ with the further decreased the ionic radius of $R$ but similar with $bc$ spiral spin order, the lower-lying peak-structure grows in intensity at 9 K and the magnitude of Im[$\epsilon\mu$] reaches the maximum about 10 [Fig. \ref{figRMnO3LightP}(c)]. The position of the lower-lying peak-structure shifts from 2.9 meV to 2 meV. On the contrary, the magnitude of Im[$\epsilon\mu$] of the higher-lying peak-structure decreases, but the peak position also shifts to the lower-energy from 7.4 meV to 5 meV; such tendency is also discerned in the Re$[\epsilon\mu]$ spectrum. The observed remarkable absorptions in TbMnO$_3$ and DyMnO$_3$ can be ascribed to the electromagnons, as discussed in detail in Sec. \ref{ExpDyMnO3}. Further systematic investigations on the variation of the optical spectra in terms of $R$ can be found in Ref. \cite{JSLee}. In the following subsections, we focus on the optical spectra at terahertz frequencies in the $A$-type AFM, $bc$ spiral, and $ab$ spiral spin ordered phases of $R$MnO$_3$, tuned by temperature and external $H$, by taking DyMnO$_3$ (Sec. \ref{ExpDyMnO3}) and Gd$_{0.7}$Tb$_{0.3}$MnO$_3$ (Sec. \ref{ExpGd07Tb03MnO3}) as examples.

\begin{figure*}[t]
\begin{minipage}[t]{.67\textwidth}
\includegraphics[width=1\textwidth]{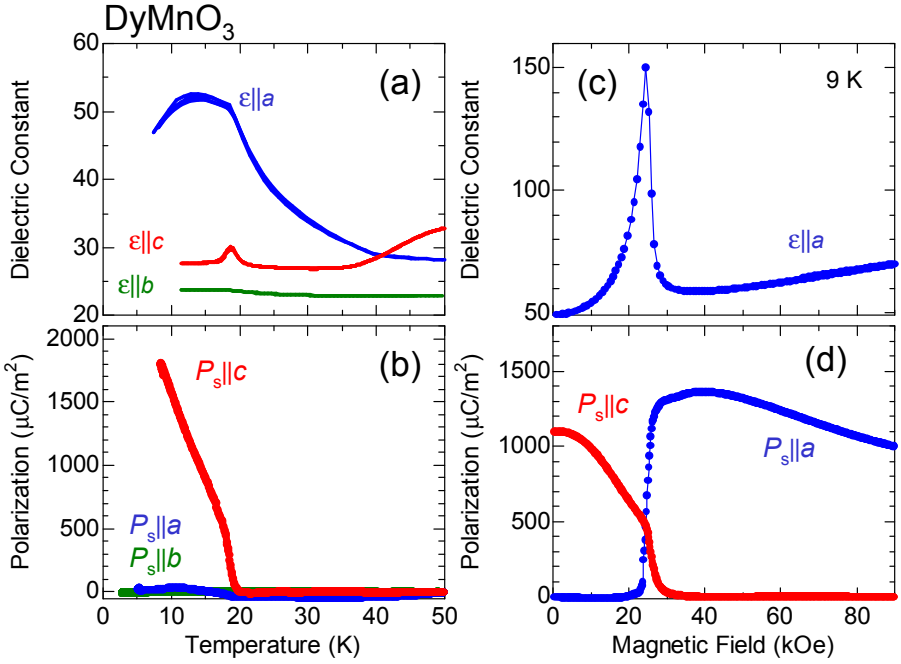}
\end{minipage}
\hfill
\begin{minipage}[b]{.3\textwidth}
\vspace*{-19cm}
\caption{(Color online) Temperature dependence of (a) dielectric constant $\epsilon$ at 10 kHz and (b) ferroelectric polarization $P_{\rm s}$ of DyMnO$_3$ with each crystallographic axis in zero magnetic field \cite{TGoto1,TKimura2}. The magnetic field $(\parallel b)$ dependence of (c) dielectric constant along the $a$-axis at 10 kHz and (d) $P_{\rm s}$ ($P_{\rm s}\| c$ and $P_{\rm s}\| a$).}
\label{DyMnO3-Ps-epsion-T}
\end{minipage}
\end{figure*}

\begin{figure*}[t]
\includegraphics[width=0.65\textwidth]{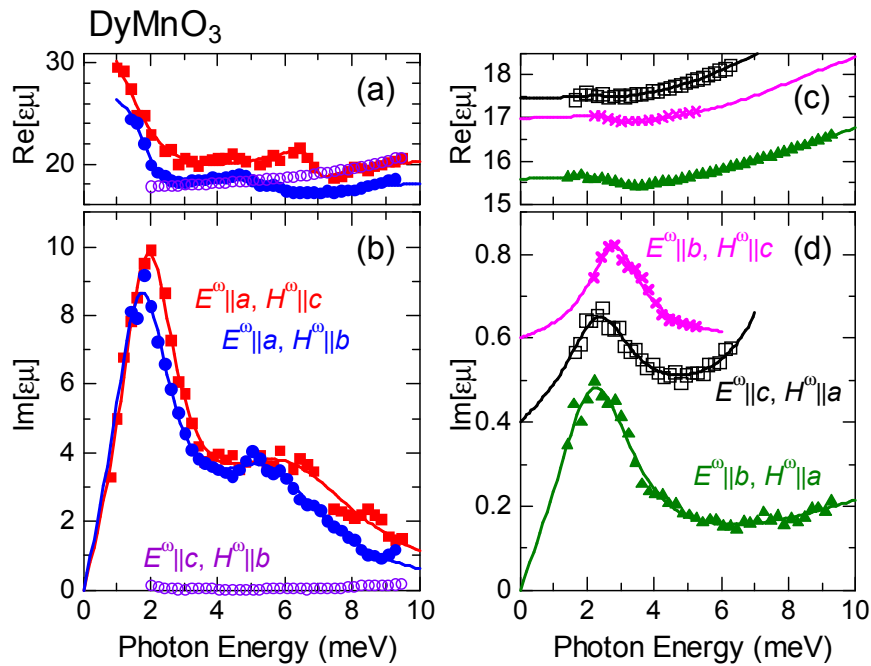}
\caption{(Color online) Light-polarization dependence of the spin excitations in the $bc$ spiral spin ordered phase of DyMnO$_3$, measured around 10 K, using a complete set of the crystal faces ($ac$, $ab$, and $bc$). The crystal orientations with respect to $E^\omega$ and $H^\omega$ are indicated in the figures. Upper and lower panels show the real Re[$\epsilon\mu$] and imaginary Im[$\epsilon\mu$] parts of the $\epsilon\mu$ spectra (symbols), respectively. Im[$\epsilon\mu]$ spectra shown in (d) are vertically offset for clarify. Note that the scales of the vertical axes in (a) and (b) are different in (c) and (d), respectively. The solid lines shown in (a) and (b) are results of a least-square fit to reproduce lower- and higher-lying peak-structures by assuming two Lorentz oscillators for $\epsilon$. On the other hand, the $\epsilon\mu$ spectra shown in (c) and (d) can be reproduced by two Lorentz oscillators for $\epsilon$ and $\mu$.}
\label{figDyMnO3LightP}
\end{figure*}

\begin{figure}[bt]
\includegraphics[width=0.43\textwidth]{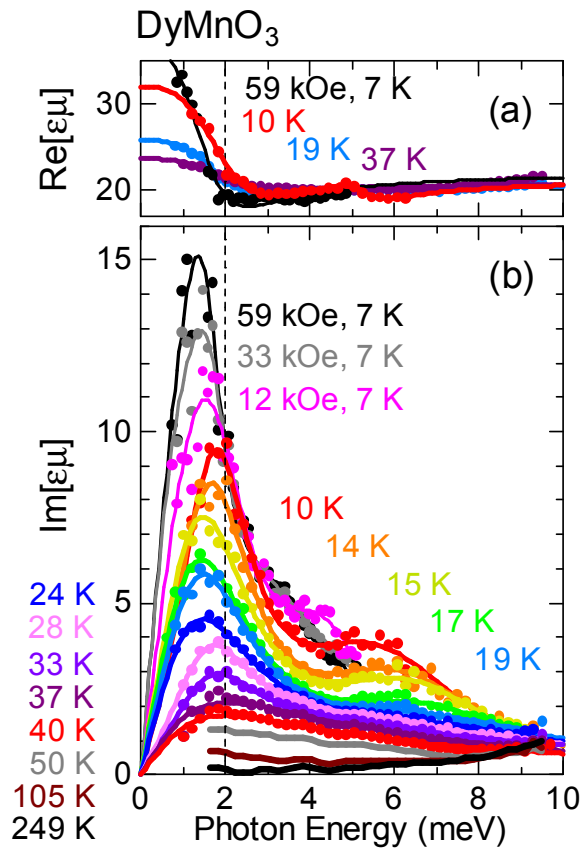}
\caption{(Color online) Temperature dependence of (a) real Re[$\epsilon\mu$] and (b) imaginary Im[$\epsilon\mu$] parts of the selected $\epsilon\mu$ spectra of DyMnO$_3$ for $E^\omega \| a$ and $H^\omega \| c$. The solid lines are results of a least-square fit to reproduce lower- and higher-lying peak-structures in the $bc$ spiral spin ordered phase below 19 K by assuming two Lorentz oscillators for $\epsilon$. The selected $\epsilon\mu$ spectra of DyMnO$_3$ in the magnetic field applied along the $b$-axis, measured at 7 K, are also included. Above 19 K, the data can be fitted by the single Lorentz oscillator.}
\label{figDyMnO3TH}
\end{figure}

\begin{figure*}[bt]
\includegraphics[width=0.88\textwidth]{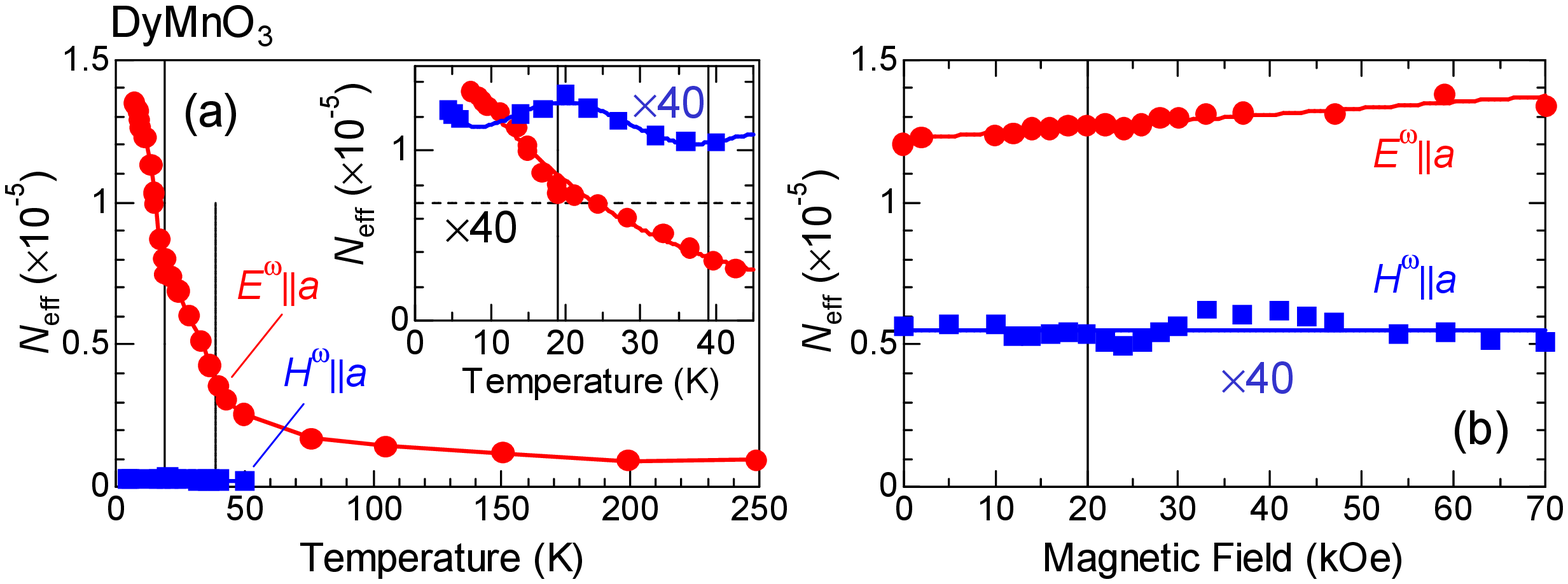}
\caption{(Color online) Integrated spectral weight per Mn-site $N_{\rm eff}$, as defined by Eq. (\ref{Neff}) in the text, of the $ac$ surface crystal plate of DyMnO$_3$ as a function of (a) temperature in zero magnetic field and (b) magnetic field at 7 K. $N_{\rm eff}$ for $E^\omega \| a$ and $H^\omega \| a$ are represented by circles and squares, respectively. The solid lines are merely the guide to the eyes. Inset of (a) shows the magnified view below 45 K. For comparison, $N_{\rm eff}$ for $H^\omega \| a$ are multiplied by 40. The horizontal dashed line in the inset of (a), also multiplied by 40, represents the estimated contribution of the background absorption for $E^\omega \| c$ to $N_{\rm eff}$ for $H^\omega \| a$. The vertical solid lines in (a) and the inset of (a) indicate the ferroelectric transition temperature of 19 K and the N\'{e}el transition temperature of 42 K. In (b), the magnetic field of 20 kOe, at which the direction of the ferroelectric polarization is flopped from the $c$- to $a$-axis or the spiral spin plane changes from $bc$ to $ab$, is indicated by the vertical solid line. }
\label{figDyMnO3Neff}
\end{figure*}

\subsection{DyMnO$_{\textbf{3}}$ with $\mbox{\boldmath$bc$}$ spiral spin order}\label{ExpDyMnO3}

In DyMnO$_3$, the collinear spin order of Mn ions evolves along the $b$-axis below $T_{\rm N}$ of 39 K. The modulation wavevector of Mn ions $q_b^{\rm Mn}$ along the $b$-axis is incommensurate with $q_b^{\rm Mn}\sim0.36$ \cite{TGoto1}. For decreasing temperature, the ferroelectricity appears along the $c$-axis as a result of the evolution of the $bc$ spiral spin order; the temperature dependence of $\epsilon$ at 10 kHz and $P_{\rm s}$ along the each crystallographic axis of DyMnO$_3$ in zero $H$ are presented in Figs. \ref{DyMnO3-Ps-epsion-T}(a) and \ref{DyMnO3-Ps-epsion-T}(b), respectively \cite{TGoto1,TKimura2}. Below the ferroelectric transition temperature $T_{\rm c}$ of 19 K, $P_{\rm s}$ along the $c$-axis steeply increases up to $\sim2000$ $\mu$C/m$^2$ [Fig. \ref{DyMnO3-Ps-epsion-T}(b)]. In accord with this, $\epsilon$ along the $c$-axis exhibits the sharp peak at $T_{\rm c}$ [Fig. \ref{DyMnO3-Ps-epsion-T}(a)]. Contrary to the behavior of $\epsilon$ along the $a$-axis of TbMnO$_3$ with the same $bc$ spiral spin order [Fig. \ref{TbMnO3-Ps-epsion-T}(a)], $\epsilon$ along the $a$-axis of DyMnO$_3$ strongly enhances below $T_{\rm N}$ and yields the large value of $\epsilon$ $(\sim50)$ at $T_{\rm c}$. $\epsilon$ along the $a$-axis is roughly by a factor of 2 larger than $\epsilon$ along other axes. We show in Fig. \ref{DyMnO3-Ps-epsion-T}(d) the $H$ dependence of $P_{\rm s}$ along the $a$- and $c$-axes, measured at 9 K. By an application of external $H$ along the $b$-axis, $P_{\rm s}$ along the $c$-axis is dramatically suppressed. On the other hand, $P_{\rm s}$ along the $a$-axis steeply increases, accompanied by the huge change of $\epsilon$ along the $a$-axis [Fig. \ref{DyMnO3-Ps-epsion-T}(c)], when the external $H$ exceeds the critical value $H_{\rm c}$ of $\sim20$ kOe. Namely, the direction of $P_{\rm s}$ can be flopped from the $c$- to $a$-axis above $H_{\rm c}$ or equivalently the spiral spin plane changes from $bc$ to $ab$. Among a family of $R$MnO$_3$, DyMnO$_3$ shows the largest ME effect, as exemplified by the remarkable change (300\% at 10 kHz) of $\epsilon$ along the $a$-axis upon the $P_{\rm s}$-flop [Fig. \ref{DyMnO3-Ps-epsion-T}(c)] and by the large $P_{\rm s}$ [Fig. \ref{DyMnO3-Ps-epsion-T}(d)] \cite{TGoto1,TKimura2}. Therefore, the terahertz time-domain spectroscopy of DyMnO$_3$ provides the useful insights into the basic characteristics of the electromagnons observed in $R$MnO$_3$ and their possible role in the ME effect.

First, we clarify the selection-rule of the spin excitation in the $bc$ spiral spin ordered phase of DyMnO$_3$ based on the light-polarization dependence (for both $E^\omega$ and $H^\omega$) using the complete set of the crystal surface plates ($ac$, $ab$, and $bc$) \cite{NKida1}. These measurements are indispensable to experimentally distinguish the respective electric and magnetic contributions to the $\epsilon\mu$ spectrum. Figures \ref{figDyMnO3LightP}(a) and \ref{figDyMnO3LightP}(b) present the Re$[\epsilon\mu]$ and Im$[\epsilon\mu]$ spectra in the $bc$ spiral spin ordered phase, respectively, measured around 10 K. There is a remarkable optical anisotropy with respect to $E^\omega$ and $H^\omega$. According to the inelastic neutron scattering experiments for the case of TbMnO$_3$ \cite{RKajimoto,DSenff1}, the crystalline-electric-field excitation of $f$ electrons lies around 4.7 meV. However, the observed remarkable optical anisotropy as well as the negligible absorption at 11 K for $E^\omega \| c$ and $H^\omega \| b$ [open circles in Figs. \ref{figDyMnO3LightP}(a) and \ref{figDyMnO3LightP}(b)] can exclude the possible emergence of such an excitation in the $\epsilon\mu$ spectrum in the measured energy range. 

In the Im$[\epsilon\mu]$ spectrum for $E^\omega \| a$ and $H^\omega \| c$ at 9 K [closed squares in Figs. \ref{figDyMnO3LightP}(a) and \ref{figDyMnO3LightP}(b)], there is a pronounced broad absorption up to 10 meV, which consists of two peak-structures around 2 meV and 6 meV. Accordingly, the clear dispersive structures are visible in the Re$[\epsilon\mu]$ spectrum. The presence of the lower-lying absorption has been reported for TbMnO$_3$ at 2.9 meV \cite{APimenov1}, GdMnO$_3$ at 2.5 meV \cite{APimenov1}, and Eu$_{1-x}$Y$_x$MnO$_3$ at 3 meV \cite{RValdesAguilar1,APimenov2}. Among them, the magnitude of Im$[\epsilon\mu]$ of DyMnO$_3$ reaches about 10, which is roughly by a factor of 2--5 larger than that of TbMnO$_3$ and GdMnO$_3$, as also seen in Fig. \ref{figRMnO3LightP}. The observed remarkable characteristics, including the positions for peak-structures and the spectral shape, can be also identified for $E^\omega \| a$ and $H^\omega \| b$ at 10 K [closed circles in Figs. \ref{figDyMnO3LightP}(a) and \ref{figDyMnO3LightP}(b)]. Therefore, the observed broad absorption can be ascribed to the electric-dipole active mode only along the $a$-axis. We can reproduce the $\epsilon\mu$ spectra for $E^\omega \| a$ by adopting two Lorentz oscillators of $\epsilon$ for peak-structures, as indicated by the solid lines. It yields Re$[\epsilon\mu(\omega\rightarrow 0)]$ of $\sim32$, which is less than $\epsilon$ along the $a$-axis at 10 kHz $(\sim50)$ [Fig. \ref{DyMnO3-Ps-epsion-T}(a)]. A noticeable characteristic is the large optical anisotropy of the Re$[\epsilon\mu]$ spectrum. Re$[\epsilon\mu]$ for $E^\omega \| a$ at low-energy yields the large value of Re$[\epsilon\mu]$ of 25--30, which is about 2 times larger than that along the other axes, i.e., 15.5--17.5. This large anisotropy ratio $(\sim2)$ is comparable to that of $\epsilon$ at 10 kHz $(\sim2)$ [Fig. \ref{DyMnO3-Ps-epsion-T}(a)], indicating that the anisotropic dielectric response is extended from kilohertz to terahertz frequencies associated the gigantic electric-dipole active excitation.

On the contrary to the $E^\omega \| a$ case, we found tiny absorptions around 2.2 meV for $H^\omega \| a$ and $H^\omega \| c$, whose peak energy is nearly identical to that of the lower-lying peak-structure for $E^\omega \| a$. Figures \ref{figDyMnO3LightP}(c) and \ref{figDyMnO3LightP}(d) show the Re$[\epsilon\mu]$ and Im$[\epsilon\mu]$ spectra, respectively, measured around 10 K. There is a single peak-structure around 2.2 meV in the Im$[\epsilon\mu]$ spectrum [Fig. \ref{figDyMnO3LightP}(d)]  with a clear dispersive structure in the Re$[\epsilon\mu]$ spectrum [Fig. \ref{figDyMnO3LightP}(c)] for $E^\omega \| b$ and $H^\omega \| a$ at 8 K (closed triangles), although the magnitudes are an order of magnitude smaller than the $E^\omega \| a$ peaks. This peak-structure can be assigned to the spin excitation driven by $H^\omega \| a$ as the nearly identical spectra signature can be observed for $E^\omega \| c$ and $H^\omega \| a$ at 6 K (open squares), while the absolute value of the $\epsilon\mu$ is slightly different. We also discern the broad peak-structure around 2.3 meV for $E^\omega \| b$ and $H^\omega \| c$ at 9 K (crosses). For the case of $E^\omega \| a$ and $H^\omega \| c$, the possible peak feature driven by $H^\omega$ is completely masked by the intense electric-dipole active absorption for $E^\omega \| a$. This peak-structure for $E^\omega \| b$ and $H^\omega \| c$ diminishes above $T_{\rm N}$, as in the same manner to the spin excitation by $H^\omega \| a$. Therefore, this peak-structure is ascribed to the spin excitation driven by $H^\omega \| c$, details of which are discussed in Sec. \ref{ExpGd07Tb03MnO3} by adopting Gd$_{0.7}$Tb$_{0.3}$MnO$_3$.

In the following, we mainly focus on the observed gigantic absorption for $E^\omega \| a$. Figures \ref{figDyMnO3TH}(a) and \ref{figDyMnO3TH}(b) show the temperature dependence of the Re$[\epsilon\mu]$ and Im$[\epsilon\mu]$ spectra, respectively, for $E^\omega \| a$ and $H^\omega \| c$. At 245 K, there is no remarkable absorption. The slight accumulation of the Im$[\epsilon\mu]$ spectrum above 6 meV is due to the contribution of the optical phonon absorption of the perovskite structure, located around 14 meV \cite{APimenov2} (see, also the overall optical spectrum of TbMnO$_3$ shown in Fig. \ref{figTbMnO3}). With decreasing temperature, the magnitude of Im$[\epsilon\mu]$ below 6 meV tends to increase. Around $T_{\rm N}$ of 42 K, Im$[\epsilon\mu]$ forms the peak-structure around 2 meV. Accordingly, there is the dispersive structure in the Re$[\epsilon\mu]$ spectrum. In the collinear spin ordered phase below $T_{\rm N}$, the absorption grows in intensity, while the peak position shifts to the lower energy. In the $bc$ spiral spin ordered phase below $T_{\rm c}$ of 19 K, the absorption becomes prominent and finally reaches the maximum of Im$[\epsilon\mu]\sim10$ at low temperature. In this ferroelectric phase, we also identify the additional peak-structure around 6 meV in the Im$[\epsilon\mu]$ spectrum, which forms the broad continuum-like absorption up to 10 meV. Its peak position shifts to the lower energy with decreasing temperature, while the lower-lying peak position shifts to the higher energy.

To further clarify the origin of the broad continuum-like absorption for $E^\omega \| a$, we show in Fig. \ref{figDyMnO3Neff}(a) the temperature dependence of the integrated spectral weight per Mn-site $(N_{\rm eff})$ defined as, 
\begin{equation}
N_{\rm eff}=\frac{2m_0V}{\pi e^2}\int_{\omega_1}^{\omega_2}\omega {\rm Im}[\epsilon(\omega)\mu(\omega)]d\omega,
\label{Neff}
\end{equation}
where $m_0$ is the free electron mass, $e$ the elementary charge, and $V$ the unit-cell volume. We chose $\omega_1=0.8$ meV and $\omega_2=9.5$ meV to fully cover the broad absorption. By the definition in Eq. (\ref{Neff}), the estimated $N_{\rm eff}$ includes the contribution of $\mu$ for $H^\omega \| c$, which is crudely estimated to be as small as 0.02 $\times$ 10$^{-5}$. Above $T_{\rm N}$, $N_{\rm eff}$ for $E^\omega \| a$ (circles) shows the negligible temperature dependence. However, $N_{\rm eff}$ gradually increases below $T_{\rm N}$ and sharply enhances below $T_{\rm c}$ [see, also in the inset of Fig. \ref{figDyMnO3Neff}(a)]. Therefore, we can conclude that the observed absorption for $E^\omega \| a$ is a magnetic in origin, namely, $E^\omega \| a$ electromagnon. On the contrary, $N_{\rm eff}$ for $H^\omega \| a$ (squares) exhibits the slight enhancement below $T_{\rm N}$ and shows the maximum at $T_{\rm c}$ [Fig. \ref{figDyMnO3Neff}(a)]. We also plot the contribution of the background absorption (a dotted line) as $N_{\rm eff}$ includes the contribution of $\epsilon$ for $E^\omega \| c$ [the inset of Fig. \ref{figDyMnO3Neff}(a)].

In $R$MnO$_3$, the rotation mode of the spiral spin plane was proposed at the early stage as the origin of the electromagnon observed at terahertz frequencies \cite{HKatsura2}. Within this picture, the rotation mode of the spiral plane has the particular selection-rule for $E^\omega$; it would become active along the $a$- and $c$-axes in the $bc$ and $ab$ spiral spin ordered phases, respectively. Such an electromagnon scenario was considered to explain the inelastic neutron scattering spectra of TbMnO$_3$ in the $bc$ spiral spin ordered phase \cite{DSenff1}. To test the origin of the electromagnon for $E^\omega \| a$, we studied the effect of $H$ on the optical properties in DyMnO$_3$ for $E^\omega \| a$ and $H^\omega \| c$. In DyMnO$_3$, applying $H$ along the $b$-axis can induce the $ab$ spiral spin ordered phase with $P_{\rm s}\| a$; the direction of $P_{\rm s}$ can be flopped from the $c$- to $a$-axis at 20 kOe, as presented in Fig. \ref{DyMnO3-Ps-epsion-T}(d). During the $P_{\rm s}$-flop, the incommensurate $q_b^{\rm Mn}$ keeps nearly constant $(\sim0.37)$ for DyMnO$_3$ \cite{JStrempfer}, contrary to the case of TbMnO$_3$, which shows the incommensurate $(q_b^{\rm Mn}=0.38)$ to commensurate $(q_b^{\rm Mn}=1/4)$ transition \cite{TArima}.

In Figs. \ref{figDyMnO3TH}(a) and \ref{figDyMnO3TH}(b), we also include the $H$ dependence of Re$[\epsilon\mu]$ and Im$[\epsilon\mu]$ spectra, respectively, measured at 7 K. The external $H$ was applied along the $b$-axis. We confirmed that the $P_{\rm s}$-flop occurs at 20 kOe [Fig. \ref{DyMnO3-Ps-epsion-T}(d)]. By application of $H$, the position of the lower-lying peak-structure shifts to the lower energy and its intensity grows. Noticeably, we can still see the electromagnon for $E^\omega \| a$ showing the gigantic absorption, even in the $ab$ spiral spin ordered phase, as exemplified by the $\epsilon\mu$ spectra at 33 kOe and 59 kOe. We plot in Fig. \ref{figDyMnO3Neff}(b) the $N_{\rm eff}$ (circles) as a function of $H$ up to 70 kOe. The integrated range of $N_{\rm eff}$ is chosen as $\omega_1=0.7$ meV and $\omega_2=5$ meV. As can be seen, there is negligible $H$ effect on $N_{\rm eff}$ even when the $ab$ spiral spin order is developed above $H_{\rm c}$ of 20 kOe apart from the slight enhancement of $N_{\rm eff}$. These observations are different from the reported result of TbMnO$_3$, in which the lower-lying peak-structure disappears by an application of the external $H$ \cite{APimenov1}. In the experiments in Ref. \cite{APimenov1}, the external $H$ was applied to the $c$-axis to destruct the ferroelectric order, accompanied by the transformation from the $bc$ spiral spin order to the $A$-type AFM order. According to the ME phase diagrams of TbMnO$_3$ and DyMnO$_3$ \cite{TKimura2}, the application of $H$ along the $b$- or $a$-axis is needed to induce the $P_{\rm s}$-flop from the $c$- to $a$-axis or equivalently the rotation of the spiral spin plane from $bc$ to $ab$, as we performed here. To confirm the survival of the electromagnon for $E^\omega \| a$, we also measured the $\epsilon\mu$ spectra for $E^\omega \| c$ and $H^\omega \| a$ in $H$ along the $b$-axis. $H$ dependence of $N_{\rm eff}$ for $E^\omega \| c$ and $H^\omega \| a$ (squares) is presented in Fig. \ref{figDyMnO3Neff}(b). We chose the integrated range as $\omega_1=2.4$ meV and $\omega_2=4$ meV. Above $H_{\rm c}$, the electromagnon arising from the rotational motion of the $ab$ spiral spin plane, is expected to appear for $E^\omega \| c$ in the $ab$ spiral spin ordered phase. However, $N_{\rm eff}$ for $E^\omega \| c$ and $H^\omega \| a$ is considerably smaller than $N_{\rm eff}$ for $E^\omega \| a$ and $H^\omega \| c$ by a factor of 100, and there is no remarkable change of $N_{\rm eff}$ at $H_{\rm c}$.

\begin{figure}[bt]
\includegraphics[width=0.46\textwidth]{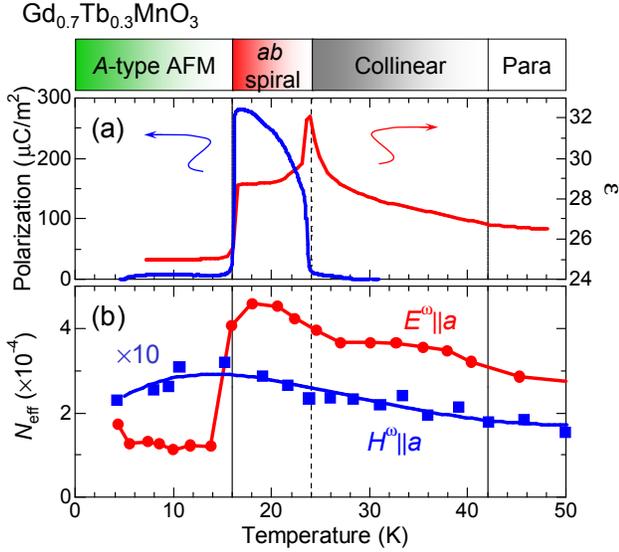}
\caption{(Color online) (a) Thermally-induced ferroelectricity in the $ab$ spiral spin ordered phase of Gd$_{0.7}$Tb$_{0.3}$MnO$_3$ \cite{TGoto2}. The ferroelectric polarization $P_{\rm s}$ emerges along the $a$-axis between 16 K and 24 K, accompanied by the change of the dielectric constant $\epsilon$ along the $a$-axis at 10 kHz. The critical temperatures for $A$-type antiferromagnetic (AFM) order with the weak ferromagnetism along the $c$-axis, $ab$ spiral spin order, and collinear spin order, are 16 K, 24 K, and 42 K, respectively, as indicated by vertical lines. (b) Temperature dependence of the integrated spectral weight per Mn-site $N_{\rm eff}$, as defined by Eq. (\ref{Neff}) in the text, of the $ac$ surface crystal plate of Gd$_{0.7}$Tb$_{0.3}$MnO$_3$. $N_{\rm eff}$ for $E^\omega \| a$ and $H^\omega \| a$ are represented by circles and squares, respectively. For comparison, $N_{\rm eff}$ for $H^\omega \| a$ are multiplied by 10. The solid lines in (b) are merely the guide to the eyes. }
\label{figGd07Tb03-Ps-ep}
\end{figure}

\subsection{Gd$_{\textbf{0.7}}$Tb$_{\textbf{0.3}}$MnO$_{\textbf{3}}$ with $\mbox{\boldmath$ab$}$ spiral spin order}\label{ExpGd07Tb03MnO3}

Here we present the results of Gd$_{0.7}$Tb$_{0.3}$MnO$_3$ in zero $H$ \cite{NKida2}, which represent the generic optical properties of the $ab$ spiral spin ordered phase of $R$MnO$_3$. Among a family of $R$MnO$_3$, Gd$_{0.7}$Tb$_{0.3}$MnO$_3$ is a rare example as it exhibits the ferroelectricity along the $a$-axis even in zero $H$ [Fig. \ref{RMnO3phasediagram}(b)]. This is in contrast to the cases of GdMnO$_3$, TbMnO$_3$, and DyMnO$_3$, where $P_{\rm s}\| a$ phase was only induced when the external $H$ is applied along the $b$- or $a$-axis, as described in Sec. \ref{ExpDyMnO3} for the case of DyMnO$_3$. We show in Fig. \ref{figGd07Tb03-Ps-ep}(a) the temperature dependences of $P_{\rm s}$ and $\epsilon$ at 10 kHz along the $a$-axis for Gd$_{0.7}$Tb$_{0.3}$MnO$_3$ \cite{TGoto2}. With decreasing temperature, the paraelectric collinear spin order evolves below $T_{\rm N}$ of 42 K along the $b$-axis, as in the similar manner to other $R$MnO$_3$ showing the $bc$ spiral spin order in zero $H$. Below $T_{\rm c}$ of 24 K, $P_{\rm s}$ steeply increases and $\epsilon$ exhibits the sharp peak. Finally, $P_{\rm s}$ reaches the maximum ($\sim280$ $\mu$C/m$^2$) at 17 K. The estimated $P_{\rm s}$ is comparable to that along the $c$-axis of TbMnO$_3$ in zero $H$ [Fig. \ref{TbMnO3-Ps-epsion-T}(b)] \cite{TKimura1,TKimura2}. Below 16 K, $P_{\rm s}$ suddenly vanishes upon the development of the $A$-type AFM order. In this $A$-type AFM phase, the finite magnetization emerges along the $c$-axis as a result of the slight canting of the Mn spins. In Gd$_{0.7}$Tb$_{0.3}$MnO$_3$, the presence of the $ab$ spiral spin ordered phase in zero $H$ was recently confirmed by the polarized neutron scattering experiments with use of an isomer of Gd ion to prevent the large cross-section of neutron of Gd ions \cite{YYamasaki2}. Moreover, $q_b^{\rm Mn}$ along the $b$-axis was found to be 0.25, which is identical to that in the $ab$ spiral spin ordered phase of TbMnO$_3$ induced by $H$ \cite{TArima}. In accord with above facts, the $ab$ spiral spin ordered phase of TbMnO$_3$ in $H$ is smoothly connected with that of Gd$_{0.7}$Tb$_{0.3}$MnO$_3$ in zero $H$, as can be seen in the comparison of ME phase diagrams for TbMnO$_3$ and Gd$_{0.7}$Tb$_{0.3}$MnO$_3$, shown in Figs. \ref{figGdTbMnO3-MEphase}(a) and \ref{figGdTbMnO3-MEphase}(b), respectively. Therefore, we can study the generic spin excitation at terahertz frequencies in the $ab$ spiral spin ordered phase, which provides the further insights into the nature of the observed electromagnon for $E^\omega \| a$ in $R$MnO$_3$.

First, we show the light-polarization dependence of the $\epsilon\mu$ spectrum for Gd$_{0.7}$Tb$_{0.3}$MnO$_3$ using a complete set of the crystal faces to clarify the light-polarization selection-rule of the spin excitation in the $ab$ spiral spin ordered phase. Figures \ref{figGd07Tb03MnO3LightP}(a) and \ref{figGd07Tb03MnO3LightP}(b) present the Re$[\epsilon\mu]$ and Im$[\epsilon\mu]$ spectra, respectively, measured around 17 K. For these measurements, we used the $ac$, $ab$, and $bc$ faces to distinguish the electric and magnetic contributions. In analogy to the $\epsilon\mu$ spectra of DyMnO$_3$ with the $bc$ spiral spin order (Fig. \ref{figDyMnO3LightP}), we observe the broad continuum-like absorption at 17 K (closed squares), which consists of two peak-structures around 2 meV and 8 meV, when $E^\omega$ and $H^\omega$ were set parallel to $a$- and $c$-axes, respectively. Such a broad absorption was assigned to the electromagnon in the $bc$ spiral spin ordered phase of DyMnO$_3$ (Sec. \ref{ExpDyMnO3}). The observed absorption of Gd$_{0.7}$Tb$_{0.3}$MnO$_3$ was also assigned to electromagnon because of the closely similar spectral feature for $E^\omega \| a$ and $H^\omega \| b$ (closed circles). Even though we checked the reproducibly of the $\epsilon\mu$ spectra by changing the thickness of the sample, the slight discrepancy of the magnitude of Im$[\epsilon\mu]$ spectrum is identified. This is perhaps due to the slight variation of Gd-to-Tb ratio as Gd$_{0.7}$Tb$_{0.3}$MnO$_3$ locates the critical region in between the paraelectric $A$-type AFM and ferroelectric $bc$ spiral spin ordered phases \cite{TGoto2}, as presented in the ME phase diagram [Fig. \ref{RMnO3phasediagram}(b)]. The magnitude of Re$[\epsilon\mu]$ for $E^\omega \| a$ at 17 K reaches about 24 at 1 meV (Fig. \ref{figGd07Tb03MnO3LightP}), which is comparable to $\epsilon$ $(\sim28)$ along the $a$-axis at 10 kHz [Fig. \ref{figGd07Tb03-Ps-ep}(a)]. The magnitude of Re$[\epsilon\mu]$ along the $a$-axis is larger than Re[$\epsilon\mu]$ for other axes by a factor of 1.5, similar to the case for DyMnO$_3$ with $bc$ spiral spin order (Fig. \ref{figDyMnO3LightP}).

A broad single peak-structure is also discerned for $E^\omega \| c$ and $H^\omega \| a$ at 17 K (open squares), where the peak energy nearly matches the peak position of the lower-lying electromagnon for $E^\omega \| a$ [Fig. \ref{figGd07Tb03MnO3LightP}(b)]. However, the magnitude of Im$[\epsilon\mu]$ at the peak energy was estimated to be 0.4, which is an order of magnitude smaller than that for $E^\omega \| a$. This absorption can be ascribed to the spin excitation driven by $H^\omega \| a$ because of the similar spectral signatures for $E^\omega \| b$ and $H^\omega \| a$ at 21 K (closed triangles). Furthermore, we observe the tiny absorption for $E^\omega \| b$ and $H^\omega \| c$ at 20 K (crosses), which becomes prominent in the $\epsilon\mu$ spectrum at 11 K in the $A$-type AFM phase, as presented in Fig. \ref{figGd07Tb03MnO3aT}(c). Upon the diminishment of the $ab$ spiral spin order, responsible for the gigantic contribution of $\epsilon$ to $\epsilon\mu$, we clearly see the single-peak-structure in the $\epsilon\mu$ spectrum at 11 K for $E^\omega \| a$ and $H^\omega \| c$. By comparing the spectrum at 11 K for $E^\omega \| a$ and $H^\omega \| c$ with that at 10 K for $E^\omega \| b$ and $H^\omega \| c$ [Fig. \ref{figGd07Tb03MnO3aT}(c)], this is attributed to the spin excitation driven by $H^\omega \| c$.

These spin excitations for $H^\omega \| a$ and $H^\omega \| c$ can be interpreted as antiferromagnetic resonances (AFMRs) of Mn ions, which are also observed in canted AFM phases of Eu$_{0.9}$Y$_{0.1}$MnO$_3$ for $H^\omega \| a$ \cite{APimenov3} and La$_{1-x}$Sr$_x$MnO$_3$ $(x<0.1)$ for $H^\omega \| c$ \cite{AAMukhin}. In fact, the $k=0$ magnon in the $A$-type AFM phase of LaMnO$_3$ locates around 3 meV \cite{KHirota}. Therefore, the measured $\epsilon\mu$ spectrum is considered to consist of a sharp resonance for $\mu$ and broad background absorption for $\epsilon$, as given by
\begin{equation}
{\rm Im}[\epsilon\mu]={\rm Re}[\epsilon]{\rm Im}[\mu]+{\rm Im}[\epsilon]{\rm Re}[\mu],
\label{epsilonmu}
\end{equation}
which can be phenomenologically expressed with two Lorentz oscillators for $\epsilon$ and $\mu$. The solid lines are the results of the least-square fit to the data using Eq. (\ref{epsilonmu}), which can reproduce the $\epsilon\mu$ spectra for Gd$_{0.7}$Tb$_{0.3}$MnO$_3$ [Figs. \ref{figGd07Tb03MnO3LightP}(c), \ref{figGd07Tb03MnO3LightP}(d), \ref{figGd07Tb03MnO3aT}(b), and \ref{figGd07Tb03MnO3aT}(c)] and DyMnO$_3$ [Figs. \ref{figDyMnO3LightP}(c) and \ref{figDyMnO3LightP}(d)].

To further see the light-polarized spectral change in a variety of the spin ordered phases, we plot in upper and lower panels of Fig. \ref{figGd07Tb03MnO3aT} the temperature variation of Re$[\epsilon\mu]$ and Im$[\epsilon\mu]$ spectra, respectively, for (a) $E^\omega \| a$ and $H^\omega \| c$, (b) $E^\omega \| c$ and $H^\omega \| a$, and (c) $E^\omega \| b$ and $H^\omega \| c$, in the $A$-type AFM phase below 16 K, ferroelectric $ab$ spiral spin ordered phase with $P_{\rm s}\| a$ between 16 K and 24 K, paraelectric collinear spin ordered phase between 24 K and 42 K, and paraelectric paramagnetic phase above 42 K.

At 10 K in the $A$-type AFM phase for $E^\omega \| c$ and $H^\omega \| a$ [Fig. \ref{figGd07Tb03MnO3aT}(b)], the clear peak-structure is identified at 2.1 meV, which can be ascribed to the AFMR driven by $H^\omega \| a$, as describe above. With increasing temperature, Im$[\epsilon\mu]$ tends to decrease in magnitude and the peak width becomes broad. In addition, the peak position shifts to the lower energy from 2.1 meV at 10 K to 1.6 meV at 33 K. Above $T_{\rm N}$ of 42 K, the peak-structure is diminished, as can be seen in the $\epsilon\mu$ spectrum at 50 K. For the case of $E^\omega \| b$ and $H^\omega \| c$ polarization [Fig. \ref{figGd07Tb03MnO3aT}(c)], a similar tendency is observed.

On the other hand, there is a remarkable temperature variation of the optical spectra for $E^\omega \| a$ and $H^\omega \| c$ [Fig. \ref{figGd07Tb03MnO3aT}(a)]. In the $A$-type AFM phase at 11 K, the sharp peak-structure can be discerned at 2.1 meV, ascribed to the conventional AFMR by $H^\omega \| c$, described above. With increasing temperature, the magnitude of Im$[\epsilon\mu]$ dramatically enhances and the electromagnon emerges as the broad continuum-like absorption, as exemplified by the $\epsilon\mu$ spectrum at 17 K in the $ab$ spiral spin ordered phase. The intensity of the electromagnon absorption decreases when the $ab$ spiral spin ordered is diminished above 24 K. However, the electromagnon survives even in the collinear spin ordered phase as a broad absorption band [lower panel of Fig. \ref{figGd07Tb03MnO3aT}(a)], the characteristic energy of which was estimated to be about 2.9 meV. Above $T_{\rm N}$ of 42 K, the broadened absorption subsists, while the intensity of the electromagnon is completely damped, as seen in the $\epsilon\mu$ spectrum at 68 K.

\begin{figure}[bt]
\includegraphics[width=0.46\textwidth]{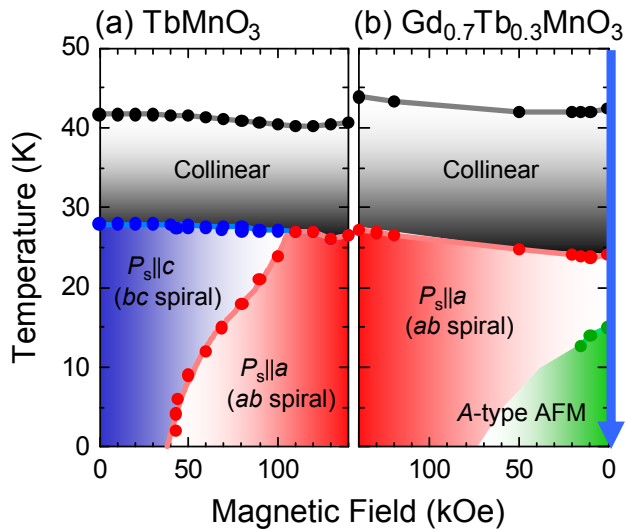}
\caption{(Color online) Magnetoelectric phase diagrams of (a) TbMnO$_3$ and (b) Gd$_{0.7}$Tb$_{0.3}$MnO$_3$. Data indicated by closed circles are taken from Refs. \cite{TKimura1,TKimura2,TGoto2} and the solid lines with shaded area are merely the guide to the eyes. The $bc$ spiral spin plane of TbMnO$_3$, acting as a source of the ferroelectric polarization $P_{\rm s}$ along the $c$-axis, can be flopped to the $ab$ spiral spin plane ($P_{\rm s}\| a$) by an application of the magnetic field $H$ along the $b$-axis. This magnetically-induced $ab$ spiral spin ordered phase smoothly connects with the thermally-induced $ab$ spiral spin ordered phase of Gd$_{0.7}$Tb$_{0.3}$MnO$_3$. The terahertz measurements in the paraelectric collinear spin ordered phase, $ab$ spiral spin ordered phase, and $A$-type antiferromagnetic (AFM) phase of Gd$_{0.7}$Tb$_{0.3}$MnO$_3$ were performed in zero $H$, as indicated by the right-side vertical arrow in (b).}
\label{figGdTbMnO3-MEphase}
\end{figure}

\begin{figure*}[t]
\includegraphics[width=0.65\textwidth]{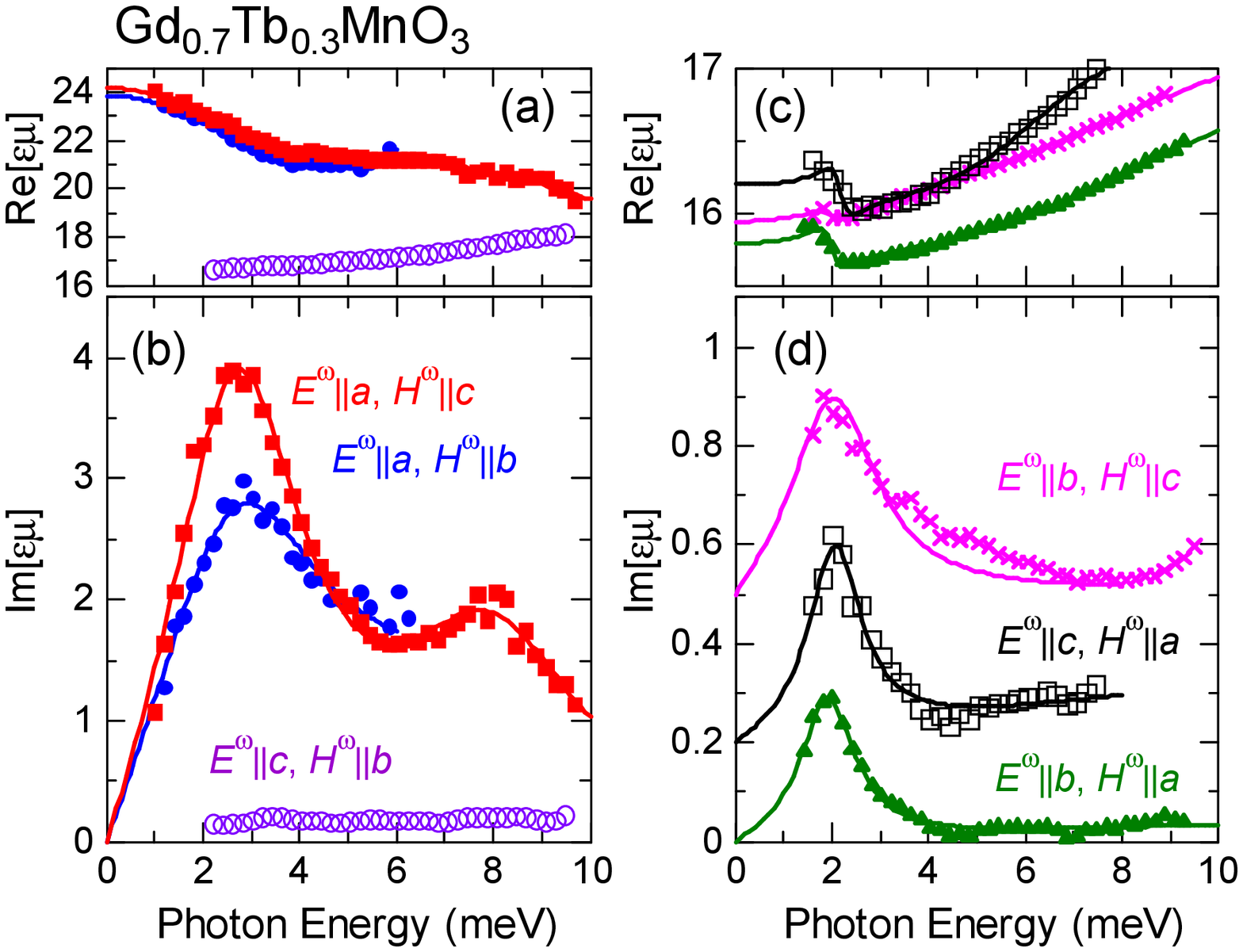}
\caption{(Color online) Light-polarization dependence of the spin excitations in the $ab$ spiral spin ordered phase of Gd$_{0.7}$Tb$_{0.3}$MnO$_3$, measured around 17 K, using a complete set of the crystal faces ($ac$, $ab$, and $bc$). The crystal orientations with respect to $E^\omega$ and $H^\omega$ are indicated in the figures. Upper and lower panels show the real Re[$\epsilon\mu$] and imaginary Im[$\epsilon\mu$] parts of the $\epsilon\mu$ spectra (symbols), respectively. Im[$\epsilon\mu]$ spectra shown in (d) are vertically offset for clarify. Note that the scales of the vertical axes in (a) and (b) are different in (c) and (d), respectively. The solid lines shown in (a) and (b) are results of a least-square fit to reproduce lower- and higher-lying peak-structures by assuming two Lorentz oscillators for $\epsilon$. On the other hand, the $\epsilon\mu$ spectra shown in (c) and (d) can be reproduced by two Lorentz oscillators for $\epsilon$ and $\mu$.}
\label{figGd07Tb03MnO3LightP}
\end{figure*}

\begin{figure*}[t]
\includegraphics[width=0.78\textwidth]{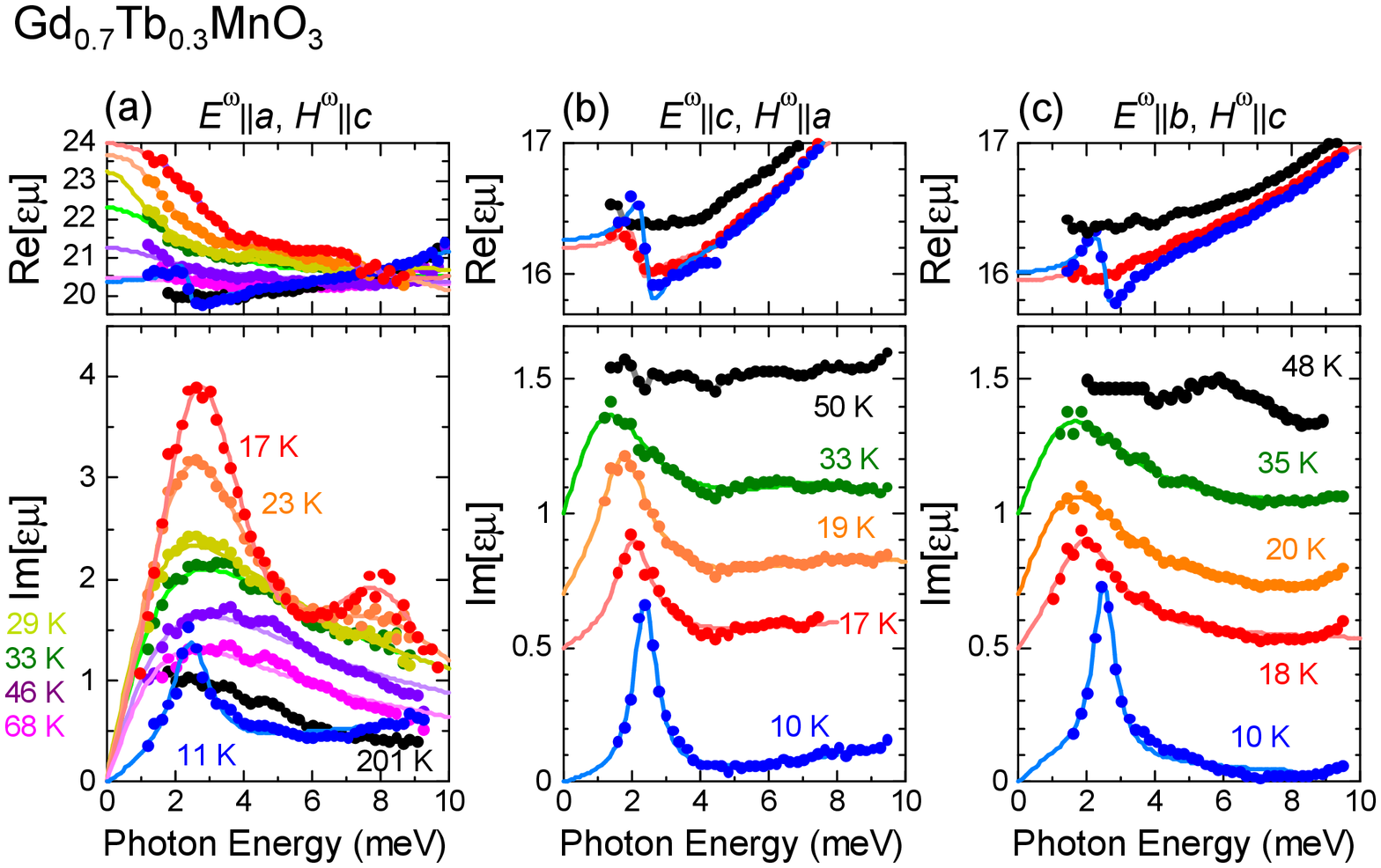}
\caption{(Color online) Temperature dependence of the real Re[$\epsilon\mu$] (upper panels) and imaginary Im[$\epsilon\mu$] (lower panels) parts of the selected $\epsilon\mu$ spectra of Gd$_{0.7}$Tb$_{0.3}$MnO$_3$ for (a) $E^\omega \| a$ and $H^\omega \| c$, (b) $E^\omega \| c$ and $H^\omega \| a$, and (c) $E^\omega \| b$ and $H^\omega \| c$. Im[$\epsilon\mu$] spectra shown in (b) and (c) are vertically offset for clarify. Note that the scale of the vertical axis in the upper panel of (a) is different in that of (b) and (c). The solid lines shown in (a) are results of a least-square fit to reproduce lower- and higher-lying peak-structures by assuming two Lorentz oscillators for $\epsilon$. The $\epsilon\mu$ spectra shown in (b) and (c) can be reproduced by two Lorentz oscillators for $\epsilon$ and $\mu$.}
\label{figGd07Tb03MnO3aT}
\end{figure*}

To be more quantitative, we show in Fig. \ref{figGd07Tb03-Ps-ep}(b) the temperature dependence of $N_{\rm eff}$ (squares) for $E^\omega \| a$; the integrated range for $N_{\rm eff}$ [see, Eq. (\ref{Neff})] was chosen from $\omega_1=1.8$ meV to $\omega_2=6.2$ meV to characterize the spin excitation spectral range. On cooling, $N_{\rm eff}$ increases below $T_{\rm N}$ and sharply enhances below $T_{\rm c}$. This behavior is in contrast to the behavior of $N_{\rm eff}$ due to the $\mu$ component AFMR with $E^\omega \| c$ and $H^\omega \| a$ (circles); the latter is nearly temperature independent above $T_{\rm N}$ and gradually increases below $T_{\rm N}$. $N_{\rm eff}$ for $E^\omega \| a$ dramatically suppresses when the $ab$ spiral spin order transforms to the $A$-type AFM order below 16 K, which contributes to the reduction of $\epsilon$ at 10 kHz [Fig. \ref{figGd07Tb03-Ps-ep}(a)].

In early studies \cite{DSenff1}, the electromagnon for $E^\omega \| a$ was ascribed to the rotation mode of the spiral spin plane. In this picture, the electromagnon would become active in the $ab$ spiral spin ordered phase of Gd$_{0.7}$Tb$_{0.3}$MnO$_3$ when $E^\omega$ was set parallel to the $c$-axis, perpendicular to the spiral spin plane. Figures \ref{figGd07Tb03MnO3LightP}(a) and \ref{figGd07Tb03MnO3LightP}(b) present Re$[\epsilon\mu]$ and Im[$\epsilon\mu]$ spectra (open circles) for $E^\omega \| c$ and $H^\omega \| b$ in the $ab$ spiral spin ordered phase, measured at 16 K. As can be seen, there is negligible absorption in the measured energy range. We also measured the $\epsilon\mu$ spectra down to 0.8 meV with use of the LT-GaAs terahertz emitters coupled with the bow-tie antenna as a light source. However, the same tendency is discerned. The present measurements using a complete set of crystal faces can clearly exclude the possibility that the observed electromagnon can be ascribed to the rotation mode of the spiral spin plane. Combined with the results presented in Sec. \ref{ExpDyMnO3} for DyMnO$_3$ in the $H$-induced $ab$ spiral spin state $(P_{\rm s}\| a)$, we can firmly conclude that there is a unique selection-rule along the $a$-axis for the electromagnons in $R$MnO$_3$, irrespective of the direction of the spiral spin plane ($bc$ or $ab$).

\section{DISCUSSION}\label{Discussion}

Based on the systematic experimental investigations with varying the ionic radius of $R$, light-polarization, temperature, and external $H$ (Sec. \ref{Results}), we can extract the general features of $R$MnO$_3$ in a variety of the spin ordered phases, as summarized below. There are spin excitations driven by $E^\omega$ and $H^\omega$ at terahertz frequencies in $R$MnO$_3$. In the ferroelectric spiral spin ordered phase, the electromagnon appears only along the $a$-axis. It spreads in the energy range of 1--10 meV as a pronounced continuum-like absorption with two peak-structures around 2 meV and 5--8 meV, rather than the single peak-structure (1--5 meV) as previously reported for TbMnO$_3$ \cite{APimenov1}. Accordingly, Re[$\epsilon\mu(\omega\rightarrow0)$] for $E^\omega$ along the $a$-axis is a factor of 2 larger than those for other polarization. This large optical anisotropy $\sim2$ is comparable to the anisotropy of $\epsilon$ at 10 kHz. The lower-lying peak-structure of the electromagnon can be identified, though broadened, even in the collinear spin ordered phase but suddenly disappears in the $A$-type AFM phase. In addition, we observed weak but sharp single peak-structure for $H^\omega \| a$ and $H^\omega \| c$ in the energy range of 1--4 meV, whose peak positions are nearly identical to that of the lower-lying electromagnon for $E^\omega \| a$. These peak-structures can be assigned to the AFMRs of Mn-ions, which are pronounced in the $A$-type AFM phase.

In the experimental data presented in Sec. \ref{Results}, we could reveal the new features of the optical properties of $R$MnO$_3$. As an origin of the electromagnon for $E^\omega \| a$, the rotation mode of the spiral spin plane has been proposed \cite{HKatsura2} based on the spin-current mechanism that can explain the emergence of ferroelectricity and the $P_{\rm s}$-flop by the external $H$ in $R$MnO$_3$ \cite{HKatsura1}. Actually, this mechanism was considered to explain the inelastic neutron scattering spectrum of TbMnO$_3$ in the $bc$ spiral spin ordered phase \cite{DSenff1}; the one of the low-energy branch of the magnon band around 2 meV was assigned to the electromagnon. Noticeably, the electromagnon model predicts the unique selection-rule in terms of the light-polarization; it becomes active for $E^\omega \| a$ and $E^\omega \| c$ in the $bc$ and $ab$ spiral spin ordered phases, respectively. However, we have clearly revealed that the electromagnon shows up only along the $a$-axis, irrespective of the direction of the spiral spin plane ($bc$ or $ab$), by measuring the $\epsilon\mu$ spectra in the $ab$ spiral spin ordered phases in DyMnO$_3$ induced by the external $H$ as well as in Gd$_{0.7}$Tb$_{0.3}$MnO$_3$ induced by temperature. There is no remarkable peak-structure in the $\epsilon\mu$ spectrum for $E^\omega \| c$ in the $ab$ spiral spin ordered phase down to 0.8 meV. Based on systematic experimental data described above, we conclude that the electromagnon in the energy range of 1--10 meV cannot be ascribed to the rotation mode of the spiral spin plane. Recently, the negligible effect of the external $H$ on the electromagnon is also confirmed in TbMnO$_3$ by applying the external $H$ along the $b$-axis \cite{RValdesAguilar2,APimenov4} or the $a$-axis \cite{APimenov4}, which is consistent with the results of DyMnO$_3$ in $H$ \cite{NKida1}.

Electromagnon can only contribute to the change of $\epsilon$ along the $a$-axis [Fig. \ref{DyMnO3-Ps-epsion-T}(a)], as evidenced by the consistency of the anisotropy ratio of Re$[\epsilon\mu]$ at terahertz and kilohertz frequencies $(\sim2)$. However, there is a still discrepancy of the absolute values of Re$[\epsilon\mu]$ at terahertz and kilohertz frequencies for the case of DyMnO$_3$, the reason of which can be understood by the presence of the ferroelectric domain wall that produces the additional contribution to $\epsilon$, as recently revealed by the dielectric spectroscopy up to 10 MHz of DyMnO$_3$ \cite{FKagawa}; the $\epsilon$ spectrum shows the relaxation type dispersion with a relaxation rate of $\sim10^{17}$ Hz and the magnitude of $\epsilon$ at 10 MHz was estimated to be $\sim35$, which seems to smoothly connect with the low-energy part of the measured Re$[\epsilon\mu]$ spectrum at terahertz frequencies (Re$[\epsilon\mu(\omega\rightarrow0)]$ $\sim$ 25--30).

As an origin of the electromagnon for $E^\omega \| a$, we proposed the conventional electric-dipole active two-magnon excitation due to the exchange-striction mechanism \cite{TMoriya}, since the observed electromagnon is independent of the direction of the spiral spin plane and spreads in the wide energy range of 1--10 meV despite the fact that the $k=0$ magnon driven by $H^\omega$ locates around 2 meV \cite{NKida1,YTakahashi1,NKida2,NKida3}. Contrary to this scenario, it was found more recently that the symmetric exchange mechanism can generate ``one-magnon" excitation when the ordered spins are non-collinear, as in the present case of cycloidal spin order \cite{RValdesAguilar2,SMiyahara}. In the following, we introduce theoretical considerations on $R$MnO$_3$ based on Heisenberg model and compare with experimental data presented in Sec. \ref{Results}.

    \begin{table}[t] 
  \caption{The nearest-neighbor ferromagnetic interaction $J_1$, next-nearest-neighbor antiferromagnetic interaction $J_2$, and the interlayer antiferromagnetic interaction $J_c$ for ${\rm DyMnO_3}$, ${\rm TbMnO_3}$, and 
      ${\rm Gd_{0.7}Tb_{0.3}MnO_3}$ \cite{SMiyahara,JSLee,DSenffR}.}
    \begin{tabular}{cccc}
      Compounds & $J_1$ (meV) & $J_2/|J_1|$ & $J_c/|J_1|$ \\ \hline
      \hline
      ${\rm DyMnO_3}$ & $-0.71$ & 1.22 & 1.5 \\
      ${\rm TbMnO_3}$ & $-0.81$ & 0.78 & 2.0 \\
      ${\rm Gd_{0.7}Tb_{0.3}MnO_3}$ & $-0.82$ & 0.75 & 2.1\\\hline
    \end{tabular}
    \label{tab:interactions}.
  \end{table}

  The magnetic behaviors of
  the cycloidal ordered magnets such as $R$MnO$_3$ can be described by 
  a three-dimensional $S = 2$ frustrated Heisenberg model:
  \begin{widetext}
    \begin{equation}
      \mathcal{H}_0 = J_1 \sum_{n.n.} \vec{S}_i \cdot \vec{S}_j 
      + J_2 \sum_{n.n.n.} \vec{S}_i \cdot \vec{S}_j
      + J_c \sum_{i.l.} \vec{S}_i \cdot \vec{S}_j
      + D_\alpha \sum_i \left( S_i^\alpha \right)^{2},
      \label{eq:Heisenberg}
    \end{equation}
  \end{widetext}
  where $J_1$ $(<0)$ is the nearest-neighbor ferromagnetic interaction, 
  and $J_2$ and $J_c$ $(>0)$, are next-nearest-neighbor and interlayer antiferromagnetic interactions, respectively (see Fig.~\ref{fig:lattice}).
  Due to the frustration between $J_1$ and $J_2$,
  spiral spin states are the ground state when
  the condition $J_2/|J_1| > 0.5$ is satisfied, where
  the spiral angle $\theta$ is given by $\cos \theta = - J_1 / 2 J_2$.
  The respective interactions for ${\rm DyMnO_3}$,  ${\rm TbMnO_3}$,  
  and ${\rm Gd_{0.7}Tb_{0.3}MnO_3}$ were estimated
  and summarized in Table.~\ref{tab:interactions}~\cite{SMiyahara,JSLee}.
  $D_\alpha$ term is a uniaxial anisotropy term, which fixes
  the direction of the spiral spin plane, i.e., 
  $\alpha = a$ for ${\rm  DyMnO_3}$ and ${\rm  TbMnO_3}$ 
  to realize a $bc$-cycloidal state
  and $\alpha = c$ for ${\rm  Gd_{0.7}Tb_{0.3}MnO_3}$ 
  to stabilize an $ab$-cycloidal state.
  We assume that $D_\alpha = 0.2 |J_1|$,
  which reproduces the anisotropy gap observed in the magnon dispersions of ${\rm TbMnO_3}$ revealed
  by inelastic neutron scattering 
  experiments.~\cite{DSenff1,SMiyahara}  

  \begin{figure}[t]
    \begin{center}
\includegraphics[width=0.48\textwidth]{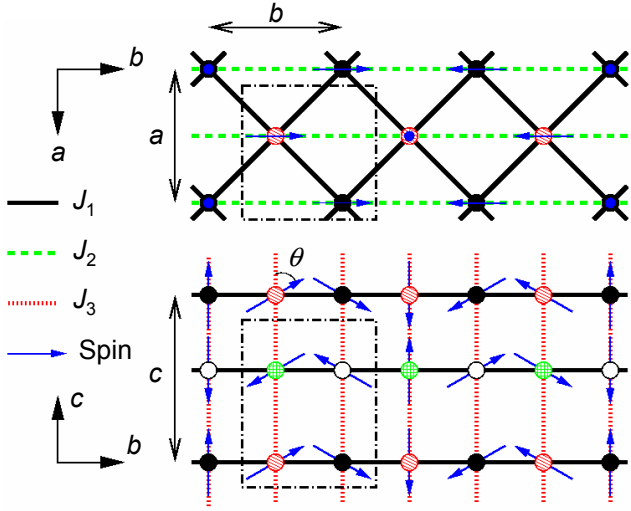}
    \end{center}
    \caption{(Color online) 
      Heisenberg model for $R{\rm MnO_3}$
      with nearest-neighbor ferromagnetic interactions $J_1$ $(<0)$,
      antiferromagnetic interactions for next-nearest-neighbor
      along the $b$-axis $J_2$ $(>0)$, and
      interlayer interactions $J_c$ $(>0)$.
      The unit cell is written by dot-dashed lines,
      where four inequivalent spin sites exist
      due to the orthorhombic lattice distortions.
      Thin (blue) arrows describe the ground state 
      spin configuration for the
      $bc$ cycloidal state.
    }
    \label{fig:lattice}
  \end{figure}

  The lower-lying magnetic excitations of
  the model on the basis of Eq.~(\ref{eq:Heisenberg}) can be described by 
  the linear spin wave theory~\cite{BRCooper,SMiyahara}.
  After rotations of local spin axis, we make Holstein-Primakoff approximations as usual,
  \begin{equation}
        S_i^+ \sim \frac{1}{\sqrt{2 S}} a_i,\;\;
    S_i^- \sim \frac{1}{\sqrt{2 S}} a_i^\dagger,\;\;
    S_i^z \sim S - a_i^\dagger a_i.
  \end{equation}
  Then, the spin Hamiltonian [Eq. (\ref{eq:Heisenberg})] is diagonalized as
  \begin{equation}
    \mathcal{H}_{0} = \sum_q \hbar \omega_q \alpha_q^{\dagger} \alpha_q
    + {\rm const}.
    \label{eq:spin-wave}
  \end{equation}
  In Eq. (\ref{eq:spin-wave}),
  the spin wave frequencies $\hbar\omega_q$ are given by
  \begin{eqnarray}
    \hbar \omega_q = 2 S \sqrt{A_q^2 - B_q^2},
  \end{eqnarray}
  where
  \begin{widetext}
    \begin{eqnarray}
      A_q & = & 2 J_{1} \cos \theta
      - \frac{J_{1}}{2} \cos^2 \frac{\theta}{2} 
      \cos \frac{\pi q_a}{a} \cos \frac{\pi q_b}{b}   
      - J_2 \cos 2 \theta + \frac{J_2}{4} \cos^2 \theta 
      \cos \frac{2 \pi q_b}{b}
      + J_c + \frac{D}{4}, \\
      B_q & = & 
      - \frac{J_{1}}{2} \sin^2 \frac{\theta}{2}  \cos \frac{\pi q_a}{a} 
      \cos \frac{\pi q_b}{b} 
      + \frac{J_2}{4} \sin^2 \theta \cos \frac{2 \pi q_b}{b} 
      + J_c \cos \frac{\pi q_c}{c} + \frac{D}{4},
    \end{eqnarray}
  \end{widetext}
  and a spin wave creation $\alpha^\dagger_q$ and annihilation $\alpha_q$ operators are
  defined as
  \begin{eqnarray}
    \alpha_q & = & c_q a_q + s_q a_{-q}^\dagger, \\ 
    \alpha^\dagger_{-q} & = & c_q a_{-q}^\dagger + s_q a_{q}, 
  \end{eqnarray}
  where $c_q$ and $s_q$ are coefficients, as given by
  $c_q  = 2 B_q/
  \{-[ 2 A_q - 2 (A_q^2 - B_q^2)^{1/2} ]^2+4 B_q^2 \}^{1/2}$, 
  and $s_q = (2 A_q - 2 (A_q^2 - B_q^2)^{1/2})
  /\{-[ 2 A_q - 2 (A_q^2 - B_q^2)^{1/2} ]^2+4 B_q^2 \}^{1/2}$.

  \begin{figure}
    \begin{center}
\includegraphics[width=0.39\textwidth]{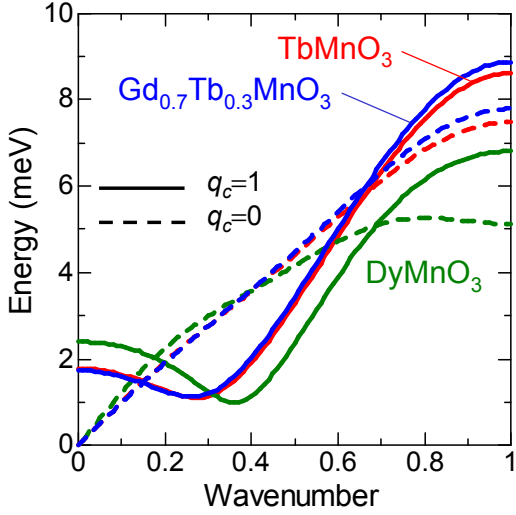}
    \end{center}
    \caption{(Color online) 
      Magnon dispersions along the $b$-axis for 
      ${\rm DyMnO_3}$, ${\rm TbMnO_3}$, and ${\rm Gd_{0.7}Tb_{0.3}MnO_3}$.
      See Table~\ref{tab:interactions} 
	for the parameter values. $q_c$ represents the wavenumber along the $c$-axis.
    }
    \label{fig:dispersions}
  \end{figure}

  \begin{figure*}
    \begin{center}
\includegraphics[width=0.88\textwidth]{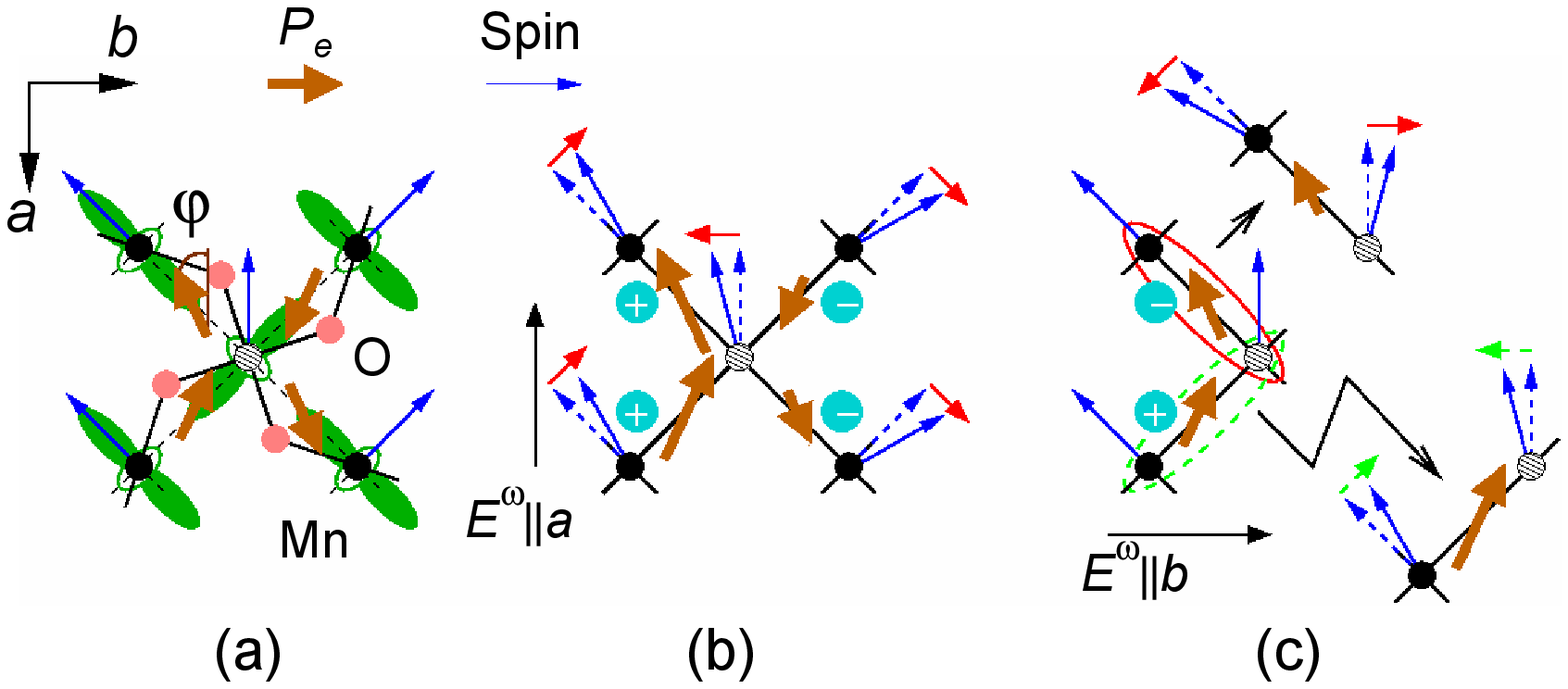}
    \end{center}
    \caption{(Color online) 
      (a) Cycloidal spin structures, orbital ordering pattern 
      on Mn sites, distortions of O sites, and direction of 
      symmetric spin-dependent polarizations $P_e$.
      The directions of $P_e$,
      which are defined by an angle $\varphi$,
      are shown by thick (brown) arrows.  
      Thin (blue) arrows
      describe the ground state spin configuration for the
      $ab$ cycloidal state.
      (b) For the case of $E^\omega \| a$ polarization, 
      the magnitudes of $P_e$ are modified, as shown on each bond due to spin structure modifications.
      Thin bold arrows are spin structures modulated by
      the electric fields and dotted arrows are spin structures 
      without electric fields. Thus oscillating electric field
      can induce the effective transverse staggered fields shown by (red) arrows,
      which can act as a source of the electric-dipole active one-magnon resonance at zone boundary.
      (c) For the case of $E^\omega \| b$ polarization, 
      the phase of $E^\omega \cdot P_e$ is staggered,
      whereas spins along the $a$-axis are uniform.
      In this case, the effective fields cancels out,
      and no one-magnon resonance occurs.
    }
    \label{fig:symmetric_EM}
  \end{figure*}

  The calculated magnon dispersions along the $b$-axis 
  are shown in Fig.~\ref{fig:dispersions}.
  Note that the uniaxial anisotropy term hardly affects
  dispersion relations
  except for $\vec{q} \sim (0, q_b^{\rm Mn}, 1)$ 
  [$q_b^{\rm Mn} \equiv \theta/(\pi/b)$].  
  The magnon dispersions calculated with 
  the parameters in Table~\ref{tab:interactions}~\cite{SMiyahara}
  can reproduce those
  observed by inelastic neutron scattering experiments
  in ${\rm TbMnO_3}$~\cite{DSenff1,DSenffR}.
  The peak positions of magnetic excitations 
  in spiral ordered phase are also estimated
  from $\hbar \omega_q$ by considering spin modes 
  for AFMRs. 
  The results are summarized in Table~\ref{tab:resonance}. 
  Here, AFMRs for 
  $H^\omega \|$ cycloidal spin plane ($H^\omega \perp$ cycloidal spin plane)
  is regarded as an excitation of the magnon at 
  $\vec{q} = (0, q_b^{\rm Mn}, 0)$ [$\vec{q} = (0, 0, 1)$].
  Note that the condition $H \perp$ cycloidal spin plane corresponds 
  to $H^\omega\| a$ for ${\rm DyMnO_3}$ and ${\rm TbMnO_3}$,
  and $H^\omega \| c$ for ${\rm Gd_{0.7}Tb_{0.3}MnO_3}$, and
  $H^\omega \|$ cycloidal spin plane
  to $H^\omega \| b$ and $H^\omega \| c$ for ${\rm DyMnO_3}$ 
  and ${\rm TbMnO_3}$, and $H^\omega \| a$ and $H^\omega \| b$ for ${\rm Gd_{0.7}Tb_{0.3}MnO_3}$.
  These magnetic resonances are observable 
  due to anisotropy like Dzyaloshinski-Moriya interactions.
  The peak positions of the AFMRs at terahertz frequencies
  driven by $H^\omega$ ~\cite{NKida1,YTakahashi1,NKida2}
  are consistent with that estimated from 
  the spin wave theory~\cite{SMiyahara}. 
  In this way, the simple Heisenberg model is a good approximation
  to describe the magnetic properties of ${\rm DyMnO_3}$,
  ${\rm TbMnO_3}$, and ${\rm Gd_{0.7}Tb_{0.3}MnO_3}$
  at low temperatures.
  Note that, to reproduce the whole complex ME phase diagram 
  of $R {\rm MnO_3}$~\cite{TKimura2}, 
  further inclusion of anisotropy terms, and Dzyaloshinski-Moriya interactions, is necessary~\cite{MMochizuki}.

  Concerning the electromagnons in $R{\rm MnO_3}$,
  it was experimentally revealed that 
  at least two modes, lower-lying mode around 2 meV 
  and the higher-lying mode around 5--8 meV, are induced 
  for $E^\omega \| a$ (Table~\ref{tab:resonance}).
  The higher-lying mode can be ascribed to the the zone boundary mode,
  i.e., magnon at $\vec{q} = (0, 1, 0)$~\cite{RValdesAguilar2,SMiyahara}.  Such a magnon is electrically induced through a symmetric term of 
  spin dependent polarizations
  $P_e = \Pi_{ij} (S_i \cdot S_j)$,
  which vanishes when a center of inversion is located
  at the middle of the bond connecting site $i$ and $j$. 
  In $R{\rm MnO_3}$, polarizations $P_e$ are realized 
  due to the $3x^2-r^2/3y^2-r^2$ orbital ordering 
  at Mn sites~\cite{SMiyahara} 
  and/or orthorhombic lattice distortions~\cite{RValdesAguilar2}, as schematically shown in Fig.~\ref{fig:symmetric_EM}(a).
  Orbital ordering induces the effective polarization
  along a bond, i.e., $\varphi = \pi/4$, and 
  distortions that perpendicular to the bond,
  i.e., $\varphi = -\pi/4$,
  where $\varphi$ is the angle from the $a$-axis. 
  Thus, in general, $\varphi$ is a free parameter in 
  $-\pi/4 \leq \varphi \leq \pi/4$, 
  however, the selection-rule for the one-magnon excitation 
  is independent of $\varphi$ as we will show in the following.
  Such an effective polarization is also obtained by
  a microscopic theory for the Mn-O-Mn bond~\cite{CJia}.
  $Pbnm$ symmetry restricts the polarization pattern, as schematically shown
  in Fig.~\ref{fig:symmetric_EM}(a).
  In such a polarization structure, 
  we can easily see that one-magnon can be induced 
  by $E^\omega$ for 
  spins with the cycloidal spin structure only for $E^\omega \| a$
  condition. On the other hand, the origin of the lower-lying mode
  is still an open question.

    \begin{table*}
  \caption{Comparison of the observed peak positions (in a unit of meV) for electromagnons and antiferromagnetic resonances (AFMRs) with theoretical estimations based on Heisenberg model for DyMnO$_3$, TbMnO$_3$, and Gd$_{0.7}$Tb$_{0.3}$MnO$_3$. Electromagnon emerges as the broad continuum-like band with two peak-structures, which become active only along the $a$-axis $(E^\omega \| a$), while AFMR appears as the single sharp peak-structure for $H^\omega \| a$ and $H^\omega \| c$.}
     \begin{tabular}{cccccccc}
      & \multicolumn{4}{c}{Experiment} & \multicolumn{3}{c}{Theory} \\
      \hline
      \hline
      Resonance & \multicolumn{2}{c}{Electromagnon} & 
      \multicolumn{2}{c}{AFMR}  & Electromagnon & 
      \multicolumn{2}{c}{AFMR}  \\
      Condition & \multicolumn{2}{c}{$E^\omega \| a$} &
      $H^\omega \| a$ & $H^\omega \| c$ &
      $E^\omega \| a$ &
      $H^\omega \| a$ & $H^\omega \| c$ \\\hline
      ${\rm DyMnO_3}$ & $\;\;\;\;$2.2$\;\;\;\;$ & 5.0 & 2.6  & 2.9 
      & 5.1  & 2.4  & 3.4  \\
      ${\rm TbMnO_3}$  & 2.9  & 7.4  & $-$ & 3.0 
      & 7.5  & 1.8  & 2.6  \\
      ${\rm Gd_{0.7}Tb_{0.3}MnO_3}$  & 2.8  & 8.0  & 2.0  & 2.0 
      & 8.0  & 2.2 & 1.6 \\\hline
    \end{tabular}
    \label{tab:resonance}
  \end{table*}

  In the external electric field $E$,
  Hamiltonian $\mathcal{H}$ can be described as $\mathcal{H} = \mathcal{H}_0 - E \cdot P_e$.
  Therefore, a spin structure is modified 
  to increase (decrease) the expectation
  values $\langle S_i \cdot S_j \rangle$,
  when $E \cdot P_e$ on the bond is positive (negative). For the spin structure in $R{\rm MnO_3}$, 
  the propagation vector of the cycloidal state 
  is aligned to the $b$-axis and spins along the $a$-axis are uniform.
  For the case of $E^\omega \| a$ stimulation, the modulation of $P_e$ is uniform along the $a$-axis, while that is staggered along the $b$-axis. Thus,
  all spins are modulated simultaneously, as schematically shown in Fig.~\ref{fig:symmetric_EM}(b),
  which can produce the effective coupling of the spin structures with $E^\omega$.
  Thus, spins are oscillated by the effective 
  transverse staggered field [Fig.~\ref{fig:symmetric_EM}(b)]
  by $E^\omega \|a$, 
  which can induce the electric-dipole active one-magnon resonance.
  On the others hand, for the case of $E^\omega \| b$ stimulation, 
  the modulation of $P_e$ along the $a$-axis
  is staggered, where the effective fields cancels out, as schematically shown in Fig.~\ref{fig:symmetric_EM}(c).
  In this way, modulated polarization does not
  couple with the spin structures
  and, thus no one-magnon resonance occurs.
  Note that these features are independent of the
  direction of the cycloidal spin plane,
  since $P_e \propto S_i \cdot S_j$.
  In this way, the observed selection-rule in $R$MnO$_3$
  is understood straightforwardly.

  These processes can be realized by representing   
  the symmetric spin dependent polarization 
  $\vec{P}_{e} = \sum S_i \cdot S_j$ by 
  using spin wave operators:
    \begin{equation}
    \sum \vec{P}_{e} \sim \vec{\Pi}(q_{zb})
    (\alpha_{q_{zb}}^{\dagger} - \alpha_{q_{zb}}),
    \label{eq:P_exchange}
  \end{equation}
  where $\vec{\Pi}(q_{zb}) = (i S \sqrt{S N}
  \Pi \cos \varphi \sin \theta (c_{q_{zb}} - s_{q_{zb}})
  , \;0,\; 0)$ and $q_{zb} \equiv (0, 1, 0)$.
  Using Eq.~(\ref{eq:P_exchange}),
  the imaginary part of complex electric polarizability tensor Im$\chi_{aa}$ at
  zero temperature, which represents an absorption, 
  is obtained from Kubo formula:
  \begin{eqnarray}
    {\rm Im} \chi_{\alpha \alpha} (\omega)
     & = & N S^3
    \Pi^2  \cos^2 \varphi \sin^2 \theta \nonumber \\
    && \;\;\;\; \times (c_{q_{zb}} - s_{q_{zb}})^2 
    \delta(\omega - \omega_{zb}) \delta_{\alpha,\, a}.
    \label{eq:chi1} 
  \end{eqnarray} 
  ${\rm Im} \chi_{\alpha \alpha} (\omega)$ has a peak at $\omega_{zb}$
  due to an electromagnon absorption.
  Such a peak-structure is consistent with the observed
  higher-lying mode around 5--8 meV in DyMnO$_3$~\cite{NKida1}, 
  TbMnO$_3$~\cite{YTakahashi1}, 
  and $\rm Gd_{0.7}Tb_{0.3}MnO_3$~\cite{NKida2} 
  on the assumption that 
  Im$[\epsilon\mu]$ $\sim$ Im$\chi_{aa}$}~\cite{SMiyahara}.

\section{SUMMARY}\label{Summary}

On the basis of the measurements of the ionic radius of $R$, light-polarization, temperature, and magnetic field dependence, we uncovered the unique optical features of the spin excitations at terahertz frequencies in multiferroic perovskite manganites, $R$MnO$_3$. We clearly identified that the electromagnon appears for $E^\omega \| a$, irrespective of the direction of the spiral spin plane ($bc$ or $ab$) or equivalently irrespective of the direction of the ferroelectric polarization ($P_{\rm s}\| c$ or $P_{\rm s}\| a$); the direct proof is provided by the spectroscopic studies on the $ab$ spiral spin ordered phases of DyMnO$_3$ induced by $H$ along the $b$-axis as well as Gd$_{0.7}$Tb$_{0.3}$MnO$_3$ induced by temperature. The observed electromagnon is broadly distributed over the measured energy range of 1--10 meV, which consists of two peak-structures around 2 meV and 5--8 meV. We also identified the $k=0$ AFMRs of Mn ions for $H^\omega \| a$ and $H^\omega \| c$ below $T_{\rm N}$ in the narrow energy range of 1--4 meV, which become prominent when the $A$-type AFM order evolves. The AFMR of Mn spins appears as the sharp peak-structure around 2 meV, which is nearly identical to the peak position of the lower-lying peak-structure of the electromagnon for $E^\omega \| a$. The electromagnon survives even in the collinear spin ordered phase, though much broadened, above the ferroelectric transition temperature, but disappears in the $A$-type AFM phase. We introduce here one of the possible scenarios to explain the observed unique light-polarization selection-rule of the electromagnon based on the Heisenberg model on the spiral spins in the orbital ordered state. With this model, the higher-lying electromagnon around 5--8 meV is assigned to the electric-dipole active one-magnon excitation at zone boundary, whose peak position agrees with the observation. However, there is a still mystery about the origin of the lower-lying electromagnon, which becomes prominent with decreasing the ionic radius of $R$. Therefore, further theoretical considerations are needed to fully explain whole spectral shape of the electromagnons at terahertz frequencies.

\subsection*{Acknowledgments}
We thank J. Fujioka and Y. Ikebe for their support to the measurements. We also thank H. Katsura, F. Kagawa, M. Mochizuki, Y. Taguchi, and N. Nagaosa for valuable discussion from an early stage of our study. This work was in part supported by Grant-In-Aids for Scientific Research (16076205 and 20340086) from the Ministry of Education, Culture, Sports, Science and Technology (MEXT), Japan.

\end{document}